\documentclass[aps,prd,showpacs]{revtex4}

\usepackage{amsmath,amsfonts}
\usepackage[dvips]{graphicx}

\newcommand{\ud}{\mathrm{d}}

\begin{document}

\title{Different faces of chaos in FRW models with scalar fields -- geometrical point of view}
\author{Orest Hrycyna}
\email{hrycyna@byk.oa.uj.edu.pl}
\affiliation{Astronomical Observatory, Jagiellonian University, Orla 171, 30-244 Krak{\'o}w, Poland}
\author{Marek Szyd{\l}owski}
\email{uoszydlo@cyf-kr.edu.pl}
\affiliation{Complex Systems Research Center, Jagiellonian University, Reymonta 4, 30-059 Krak{\'o}w, Poland}

\begin{abstract}
FRW cosmologies with conformally coupled scalar fields are investigated in a geometrical way by the means of geodesics of the Jacobi metric. In this model of dynamics, trajectories in the configuration space are represented by geodesics. Because of the singular nature of the Jacobi metric on the boundary set $\partial\mathcal{D}$ of the domain of admissible motion, the geodesics change the cone sectors several times (or an infinite number of times) in the neighborhood of the singular set $\partial\mathcal{D}$.

We show that this singular set contains interesting information about the dynamical complexity of the model. Firstly, this set can be used as a Poincar{\'e} surface for construction of Poincar{\'e} sections, and the trajectories then have the recurrence property. We also investigate the distribution of the intersection points. Secondly, the full classification of periodic orbits in the configuration space is performed and existence of UPO is demonstrated. Our general conclusion is that, although the presented model leads to several complications, like divergence of curvature invariants as a measure of sensitive dependence on initial conditions, some global results can be obtained and some additional physical insight is gained from using the conformal Jacobi metric. We also study the complex behavior of trajectories in terms of symbolic dynamics.
\end{abstract}


\maketitle

\section{Introduction}

In the present paper we work in a certain geometrical model of dynamics which can be constructed due to some variational principles of mechanics that allow one to construct an isomorphism between the dynamics and its model.

Recently a geometric description of chaos in Hamiltonian systems of cosmological origin has been formulated using the tools of Riemannian (or pseudo-Riemannian) geometry \cite{Pettini:1993,Casetti:2000}. We concentrate on the approach to dynamics of the systems with a natural Lagrangian function $\mathcal{L} = \frac{1}{2} g_{\alpha \beta} \dot{q}^{\alpha} \dot{q}^{\beta} - V(q)$. On the level of constant energy $E$ this system can be reduced to the geodesics flow on a pseudo-Riemannian manifold with the Jacobi metric $\hat{g}_{\alpha \beta}=2(E-V)g_{\alpha \beta}$ \cite{Szydlowski:1996a,Szydlowski:1993a,Szydlowski:1993b,Szydlowski:1990}. The conformal metric becomes degenerate for certain values of energy $E$, at some points of configuration space $\{E=V\}$ for classical systems as well as for general relativistic ones. As a consequence one has the complication of matching together geodesics well defined on open sets across a singular surface and the divergence of curvature invariants characterizing the property of sensitive dependence on initial conditions. One can avoid this crucial problem addressed in the context of Mixmaster models \cite{Burd:1993,Motter:2001} by formulating corresponding dynamical system as a system in the Finsler \cite{Cipriani:1998} or Eisenhart metric \cite{Rama:2001}.

The fact that Jacobi geometry is singular suggests that this model of dynamics \footnote{For interesting applications of Jacobi metric in the context of Schr{\"o}dinger equation see \cite{Faraoni:2002}} is one of the worst possible choices when it comes to achieving characterization of property of sensitive dependence on initial conditions it terms of curvature invariants or when it comes to achieving global results. It will be demonstrated that it is not true for the example of the behavior of geodesics in the Jacobi metric for FRW cosmological models with conformally coupled scalar field. Of course, because of the singular nature of the Jacobi geometry, the geodesics change from time-like to space-like (sometimes several or an infinite number of times) during their evolution, but in order to ``piece together the results'', one is forced to try to match geodesics (which are representing periodic orbits of their original dynamics) each time the solution passes a singular point. This leads to the full classification of periodic orbits in terms of ``just piecing together'' segments of geodesics. It is an example of the fact that some global theorem about geodesics can be achieved. Using this classification numerous methods based on symbolic dynamics and detection of the existence of unstable periodic orbits are proposed toward an invariant characterization of dynamical complexity in cosmology \cite{Motter:2003}.

The organization of the paper is the following. In section~\ref{sec:2} the cosmological FRW model with conformally coupled scalar field is presented as a Hamiltonian dynamical system. This system belongs to larger class of simple indefinite dynamical systems for which kinetic energy form is quadric in the momenta and has Lorentzian signature (is indefinite). In section~\ref{sec:3} we consider Poincar{\'e} sections of the trajectories, with the surface formed by the boundary set of the domain classically admissible for motion, as a indicator of complex behavior. Note that the trajectories have property of recurrence due to multiple intersections of singular set. In this section the full classification of simple periodic orbits is performed and the largest Lyapunov exponent is numerically calculated. Section~\ref{sec:3} also contains the analysis of distributions of intersection points on the singular set. The existence of a weak noise observed in the Fourier analysis of intersections seems to be a complementary indicator of chaotic behavior.

\section{FRW models with scalar fields as a simple indefinite dynamical systems}
\label{sec:2}

The dynamical systems of cosmological origin have many special features which distinguish them from those met in classical mechanics. It is in fact the origin of some problems and controversy in understanding of chaos in cosmology \cite{Motter:2002,Castagnino:2001}.

In this section we shall study the dynamical complexity of a simple inflationary models of the Universe, regarded as Hamiltonian dynamical systems. They appeared, for example, in Linde's chaotic inflation scenario, where inflation is driven by the vacuum energy of a single slowly rolling inflaton field \cite{Linde:1983}.

The dynamics of cosmological models, allowing for an inflaton field, has been studied by many authors \cite{Belinsky:1955,Belinsky:1985}. The idea of inflation which was introduced to solve some of the underlying problems in the standard big-bang cosmology becomes strictly connected with the existence of a scalar field which generates the period of an accelerated expansion of the Universe. In this case its energy density becomes dominated by the potential energy $V(\phi)$ of the scalar field $\phi$ (the inflaton). Although the dynamics of inflation depends on the specifics of the models, the basic mechanism lies in the equation of motion which for a homogeneous scalar field ($\phi=\phi(t)$) takes the form
\begin{equation}
\ddot{\phi} + 3 H \dot{\phi} + V'(\phi) + \frac{R}{6} \phi = 0 ,
\label{eq:1}
\end{equation}
where an over dot represents a derivative whit respect to the cosmological time $t$, and $V'=\ud V/\ud\phi$, $R$ is the Ricci scalar here, and the last term vanishes for minimally coupled scalar fields (in general the last term is $\xi R \phi$, where $\xi=1/6$ for the case of conformally coupled scalar fields).

Here we deal with a single homogeneous and conformally coupled scalar field $\phi$ with potential $V(\phi)$ on the FRW background. The same system was previously considered in terms of the original dynamical systems without using the conformal Jacobi metric by many authors in the context of inflation and chaos \cite{Blanco:1995,Calzetta:1993,Cornish:1996}. Our goal in this paper is to point out different manifestations of the complex dynamics in terms of geodesics in the Jacobi metric. The motivation for such a study is to obtain an additional physical insight using the conformal metric and the behavior of periodic orbits.

Our cosmological model assumes the FRW spatially geometry, that is, the line element is of the form
\begin{equation}
\ud s^{2} = a^{2}(\eta) \{ - \ud\eta^{2} + \ud\chi^{2} +f^{2}(\chi)(\ud\theta^{2} + \sin^{2}{\theta} \ud\varphi^2)\},
\label{eq:2}
\end{equation}
where $0 \le \varphi \le 2\pi$, $0 \le \theta \le \pi$, $0 \le \chi \le \pi$ and $\eta$ is the conformal time $\ud t/a=\ud\eta$; a -- the scale factor;
\begin{equation}
f(\chi) = \left \{
\begin{array}{lll}
\sin{\chi},  & 0 \le \chi \le \pi    & k=+1 \\
\chi,	     & 0 \le \chi \le \infty & k=0 \\
\sinh{\chi}, & 0 \le \chi \le \infty & k=-1
\end{array}
\right.
\nonumber
\end{equation}
$k$ is the curvature index.

The gravitational dynamics is derived from the Einstein-Hilbert action
\begin{equation}
S_{g} = \frac{1}{2} \int \ud^{4} x \sqrt{-g}(R - 2 \Lambda),
\label{eq:3}
\end{equation}
where $g$ is the determinant of the metric; $-g = a^{4} f^{2}(\chi) \sin{\theta}$ and $R$ is the curvature scalar for metric (\ref{eq:2}) given by
\begin{equation}
R = 6 \bigg\{ \frac{\ddot{a}}{a^{3}} + \frac{k}{a^2} \bigg\},
\label{eq:4}
\end{equation}
where a dot denotes differentiation with respect to $\eta$. We use the Misner-Thorne-Wheeler (MTW) convention \cite{Misner:1972} as well as a natural system of units, such that $\hbar = c = 8 \pi G = 1$ and $(2 \pi)^{2} = 1$. In so far as Robertson-Walker symmetry holds, the scalar field should be homogeneous $\phi=\phi(t)$.

The action for a conformally coupled (real) scalar field is given by
\begin{equation}
S_{\phi} = -\frac{1}{2} \int \ud^{4}x \sqrt{-g} \big\{ \partial_{\mu}\phi \partial^{\mu} \phi + \frac{1}{6} R \phi^{2} - 2 V(\phi) \big\} 
\label{eq:5}
\end{equation}
where $V(\phi) = \frac{1}{2} m^2 \phi^{2} + \frac{\lambda}{4} \phi^{4}$ is the assumed form of potential for this scalar field.

After integration over the spatial variables ($\int \ud^{3} x = 2\pi^{2}$ is the conformal volume of an spatial hypersurface of constant curvature) and discarding total derivatives in the full action we obtain dynamical system with two degrees of freedom: $a$ and a rescaled scalar field $\psi : \phi \to \psi = \sqrt{1/6} a \phi$ with Hamiltonian
\begin{equation}
\mathcal{H} = \frac{1}{2} \{ -(p_{a}^{2} +k a^{2}) + (p_{\psi}^{2} + k \psi^{2}) + m^{2} a^{2} \psi^{2} + \lambda \psi^{4} + \Lambda a^{4} \},
\label{eq:6}
\end{equation}
where $m^{2}$ is the mass of the scalar field, $\Lambda$ is constant proportional by the factor of $1/3$ to the cosmological constant.

The evolution of the system should by considered on the $\mathcal{H}=0$ energy level for vacuum cosmology or on the $\mathcal{H}= -\rho_{r,0}$ energy surface if we add a radiation component to the energy-momentum tensor whose energy density scales like $\rho_{r}=\rho_{r,0} a^{-4}$, where $\rho_{r,0}=$ const \cite{Joras:2003}. Let us note that system (\ref{eq:6}) belongs to a larger class of dynamical systems which we call simple indefinite mechanical systems.

By simple indefinite mechanical system we understand the triple $(\mathcal{M}, g, V)$, where $\mathcal{M}$ is the configuration space carrying a metric $g$ which defines the indefinite kinetic energy form $\mathcal{K}=\frac{1}{2}g(\mathbf{u},\mathbf{u})$, $\mathbf{u} \in T_{x}\mathcal{M}$, $x \in \mathcal{M}$. $V$ is the potential function $V : \mathcal{M} \to \mathcal{R}$ which is $C^{\infty}$, and $g$ has the Lorentz signature $(-,+,+,+)$.

The ``simple'' in the above context means that dynamical system has the natural form of Lagrange function $\mathcal{L} = \frac{1}{2} g_{\alpha \beta} \dot{q}^{\alpha} \dot{q}^{\beta} - V(q)$, where $\alpha , \beta = 1,\ldots,N$ and $q$ and $\dot{q}$ are generalized coordinates and velocities respectively. The Hamilton function for such a system is of the form
\begin{equation}
\mathcal{H}(p,q) = \frac{1}{2} g^{\alpha \beta} p_{\alpha} p_{\beta} + V(q), \qquad p_{\alpha} = g_{\alpha \beta} \dot{q}^{\beta}.
\label{eq:7}
\end{equation}
For the system under consideration $(q^{1},q^{2})=(a,\psi)$. Because of our general relativity and cosmology application $\mathcal{H}=E=$ const $\iff g_{\alpha \beta} \dot{q}^{\alpha} \dot{q}^{\beta} = 2(E-V(q))$. Therefore trajectories of the system in the tangent space of $\mathcal{R}^{2N}$ with coordinates $(q^{\alpha},\dot{q}^{\alpha})$ are situated in the domain described by $\Omega = \{(q^{\alpha},\dot{q}^{\alpha}) \in \mathcal{R}^{2N} : g_{\alpha \beta} \dot{q}^{\alpha} \dot{q}^{\beta} = \| \mathbf{u} \|^{2} = 2(E-V(q)) \}$.

In the tangent space $T_{q}(\mathcal{R}^{N})$ it is natural to distinguish three classes of vectors, namely, a vector $\mathbf{u}$ is time-like if $\|\mathbf{u}\|^{2} < 0$, space-like if $\|\mathbf{u}\|^{2} > 0$ and null if $\|\mathbf{u}\|^{2}=0$.

In the configuration space we distinguish three subsets
\begin{equation}
\begin{array}{l}
\mathcal{D}_{S} = \{q\in\mathcal{R}^{N} : E-V(q) < 0 \}, \\
\mathcal{D}_{T} = \{q\in\mathcal{R}^{N} : E-V(q) > 0 \}, \\
\partial\mathcal{D} = \{q\in\mathcal{R}^{N} : E-V(q) = 0 \}.
\end{array}
\label{eq:8}
\end{equation}
Note that set $\partial\mathcal{D}$ is the boundary set because in its neighborhood we can always find points of $\mathcal{D}_{S}$ and $\mathcal{D}_{T}$
\begin{equation}
\partial\mathcal{D}_{S} = \partial\mathcal{D}_{T} = \partial\mathcal{D}.
\label{eq:9}
\end{equation}

In the three distinguished domains the character of the vector tangent to a trajectory is determined by the Hamiltonian constraint $\mathcal{H}=E$. Therefore if a trajectory changes the domain, say $\mathcal{D}_{S}$ to $\mathcal{D}_{T}$, it crosses $\partial\mathcal{D}$ and the tangent vector to that trajectory at the point $q \in \partial\mathcal{D}$ is situated on the cone determined by the kinetic form: $g_{\alpha \beta} \dot{q}^{\alpha} \dot{q}^{\beta} = E$. In Ref. \cite{Burd:1993} we can find the proof that in the case of B(IX) model trajectory crosses the boundary set $V=0$. During each of oscillations that occur close to boundary set the solution is instantaneously Kasner and therefore $V=0$ an infinite number of times.

\begin{figure}[t]
\begin{center}
\includegraphics[scale=0.65]{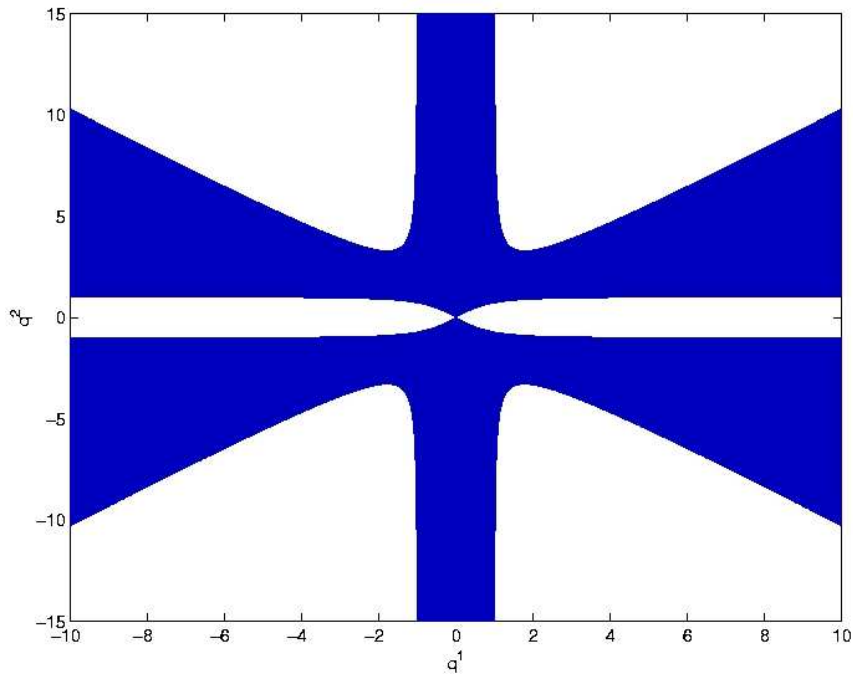}
\end{center}
\caption{The domain admissible for motion of the FRW closed system with a conformally coupled scalar field. In the shaded area trajectories behave locally unstable $K g(\mathbf{u},\mathbf{u}) < 0$ where $K$ is the Gauss curvature for the Jacobi metric.}
\label{fig:1}
\end{figure}

The physical trajectories of the simple indefinite systems are geodesics on pseudo-Riemannian manifold without the boundary (on which metric is degenerate) if we define the metric
\begin{equation}
\hat{g}_{\alpha \beta} = 2 | E-V | g_{\alpha \beta}
\nonumber
\end{equation}
and reparameterize the time variable $\eta \to s$ \cite{Szydlowski:1996b}:
\begin{equation}
\frac{\ud s}{\ud \eta} = 2 | E-V |.
\nonumber
\end{equation}
In our further considerations, we will demonstrate the advantages of analysis behavior of trajectories in a neighborhood of the boundary set $\partial\mathcal{D}$ of the region classically admissible for motion. It is also a surface of degeneration of the Jacobi metric. This surface and points of its intersections with the trajectories of the system contain interesting information about complex behavior. For simplicity of presentation of this idea we assume $k=+1$, i.e. vacuum and closed FRW model is considered. The domain (classically) admissible for motion, as well as the boundary set, are shown on Fig. \ref{fig:1} ($q^{1}=a$, $q^{2}=\psi$).

\section{Different evidences of chaotic behavior}
\label{sec:3}

\subsection{Poincar{\'e} sections}

It is clear that if a trajectory passes trough the boundary set $\partial\mathcal{D}$, then tangent vector to the trajectory is null, i.e. it lies on the cone $g_{\alpha \beta} \dot{q}^{\alpha} \dot{q}^{\beta}=0$. The physical trajectories of the indefinite mechanical systems for given total energy $E$ (zero for vacuum cosmology) are geodesics if we choose metric in the form $\hat{g}_{\alpha \beta} = 2(E - V) g_{\alpha \beta}$ in both domains $\mathcal{D}_{S}$ and $\mathcal{D}_{T}$. On the boundary $E=V$ the metric is degenerate, which is a source of obstacles if we define the property of sensitive dependence on initial conditions in terms of curvature invariants (the Gauss curvature in our case). In both open regions $\mathcal{D}_{S}$ and $\mathcal{D}_{T}$ the Euler-Lagrange equation for the Lagrangian $\mathcal{L}= \frac{1}{2} g_{\alpha \beta} \dot{q}^{\alpha} \dot{q}^{\beta} - V(q)$, ( $\dot{ }=\frac{\ud}{\ud t}$) assumes the form of geodesics equation after reparameterization of the time variable \cite{Szydlowski:1996b}
\begin{equation}
t \to s : \frac{\ud s}{\ud t} = 2 | E-V |
\label{eq:17}
\end{equation}
where $s$ is the natural parameter defined along geodesics.

The criterion of local instability of geodesics flow can be formulated as $K g(\mathbf{u},\mathbf{u}) < 0$, where $\mathbf{u}$ is the vector tangent to the trajectory and $K$ is the Gauss curvature for the Jacobi metric. This domain is represented by shaded regions on Fig. \ref{fig:1}. Due to the representation dynamics as a geodesics flow one can imagine fictitious free falling particle in both domains $\mathcal{D}_{S}$ and $\mathcal{D}_{T}$ meeting the singularity (infinite curvature $K$) at the boundary set $\partial\mathcal{D}$ which play the role of a scattering surface.

\begin{figure}
\begin{center}
\includegraphics[scale = 0.45]{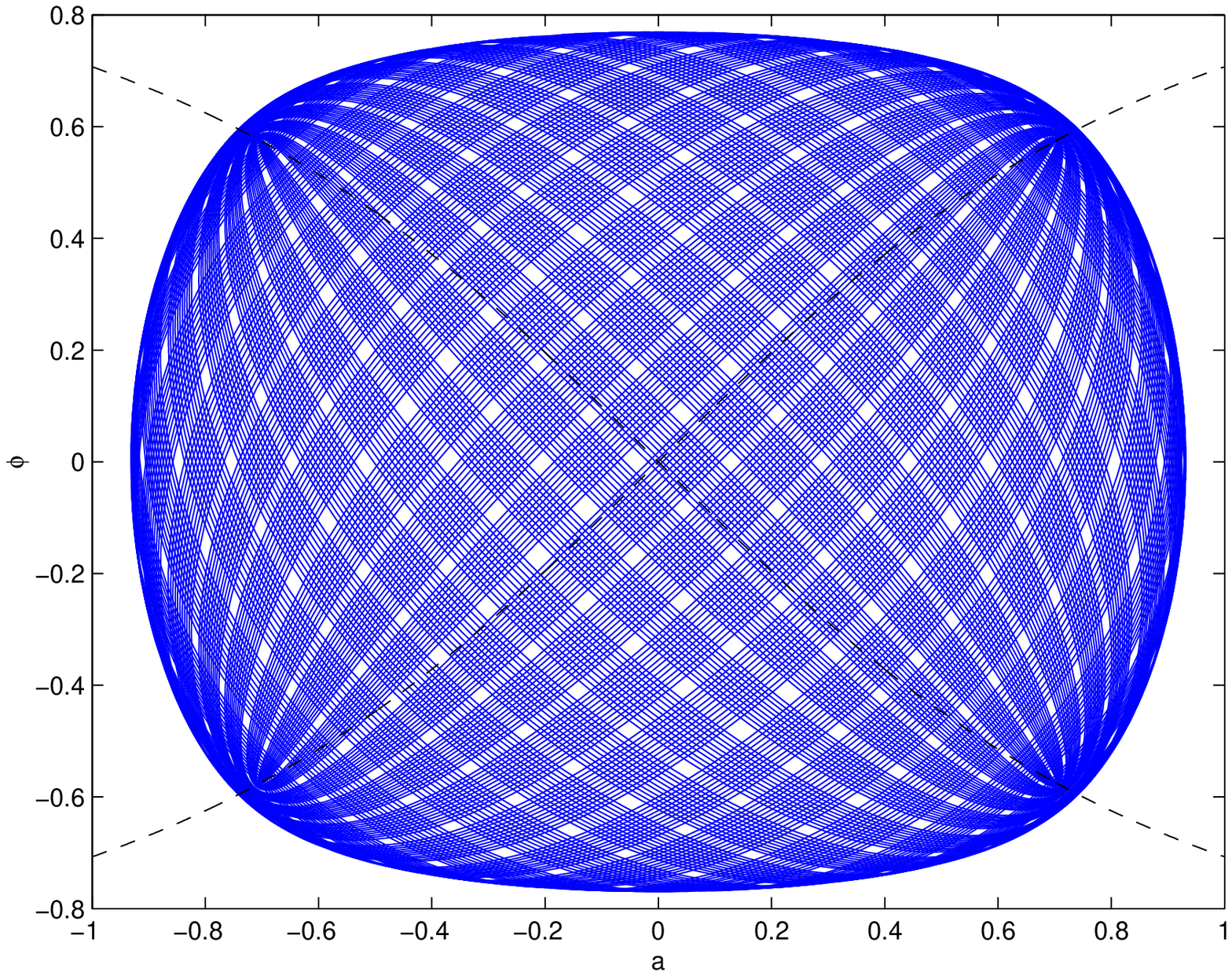}
\includegraphics[scale = 0.45]{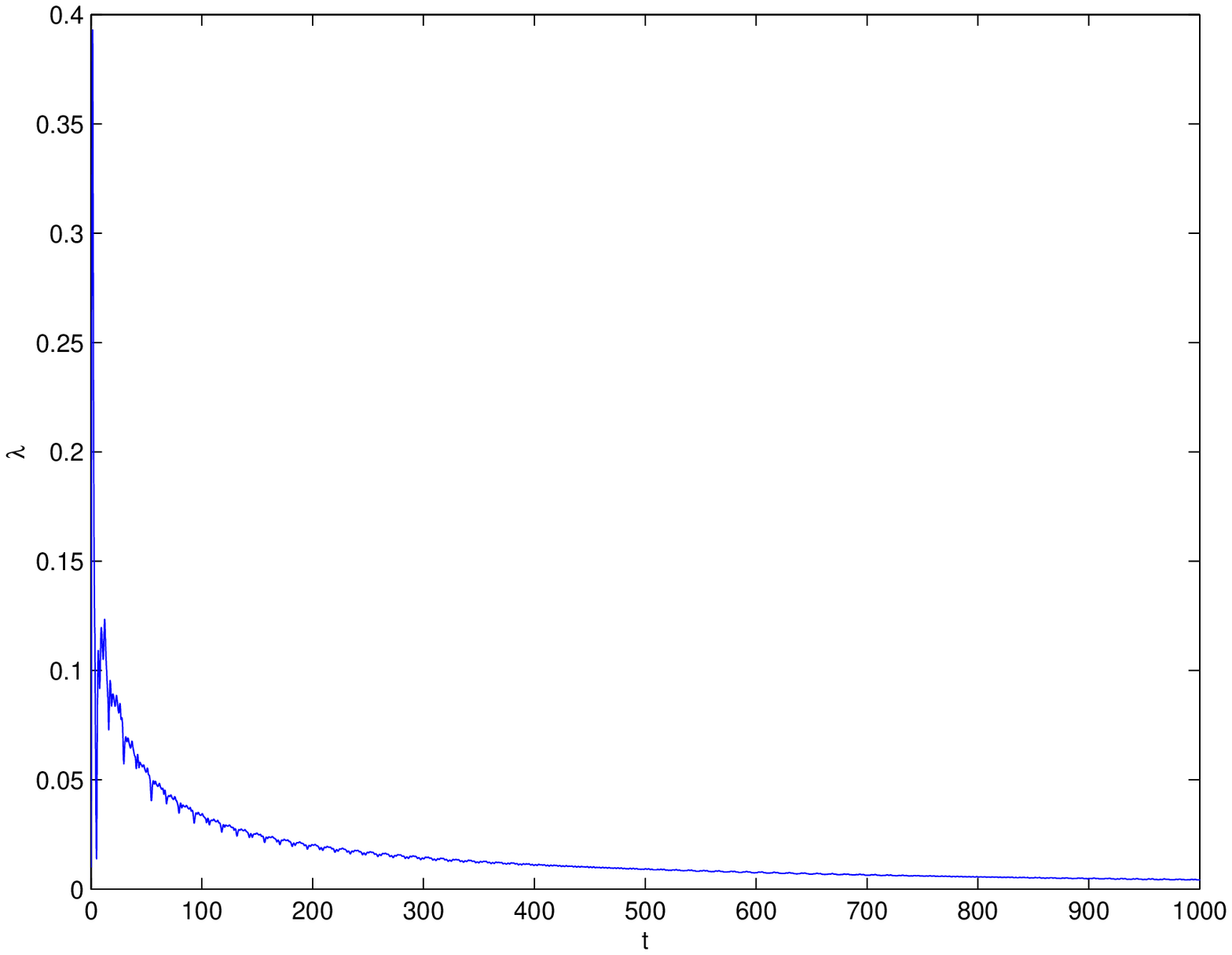}
\caption{The trajectory in the configuration space for initial conditions $a_{0}=0$, $\phi_{0}=\dot{\phi}_{0}=0.5$ ($\dot{a_{0}}$ calculated from Hamiltonian constraint) and Lyapunov principal exponent which tends to zero which may suggest that the system is regular.}
\label{fig:2}
\end{center}
\end{figure}

\begin{figure}
\begin{center}
\includegraphics[scale = 0.45]{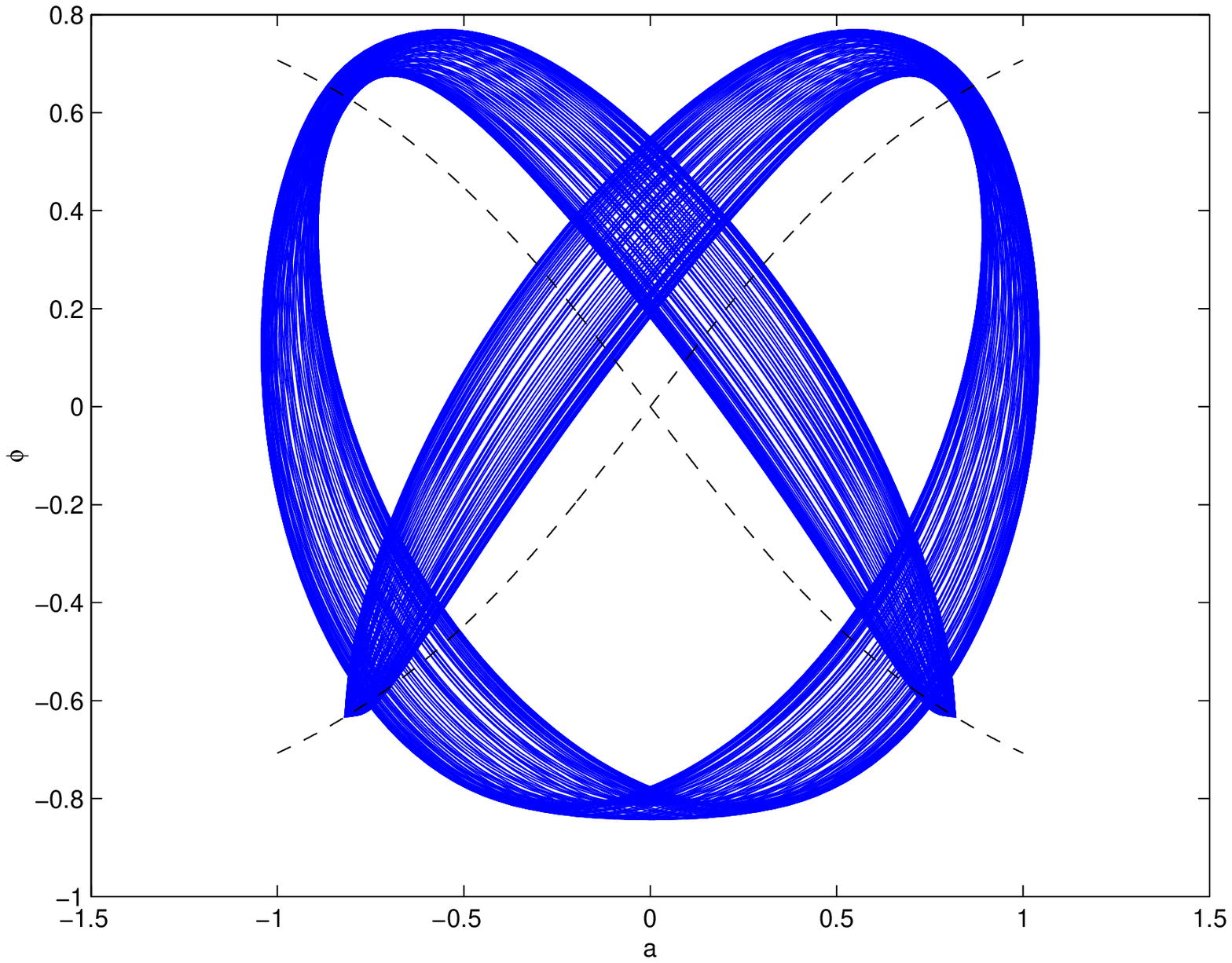}
\includegraphics[scale = 0.45]{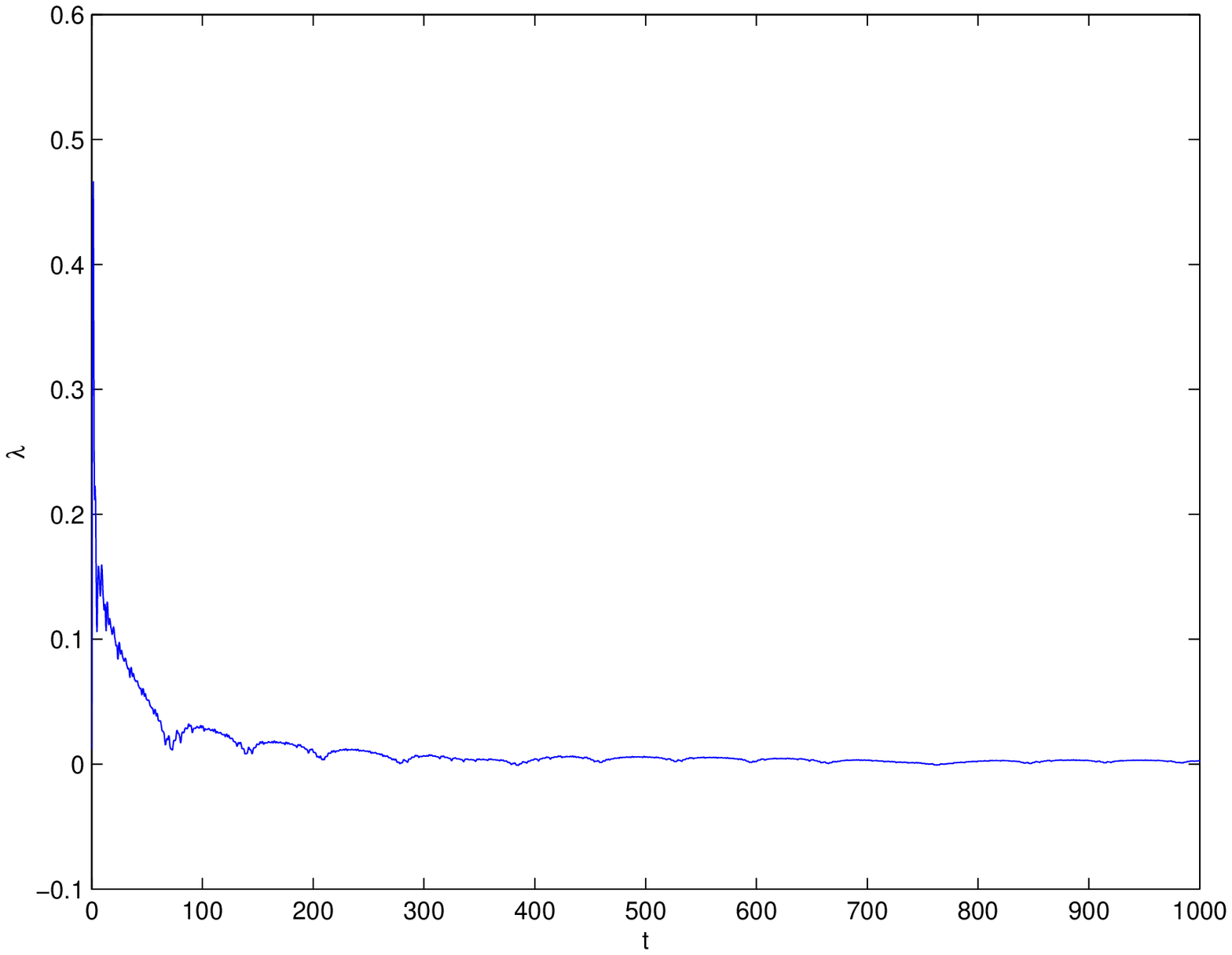}
\caption{The trajectory in the configuration space for initial conditions $a_{0}=0$, $\phi_{0}=\dot{\phi}_{0}=0.55$ ($\dot{a_{0}}$ calculated from Hamiltonian constraint) and Lyapunov principal exponent.}
\label{fig:3}
\end{center}
\end{figure}

\begin{figure}
\begin{center}
\includegraphics[scale = 0.45]{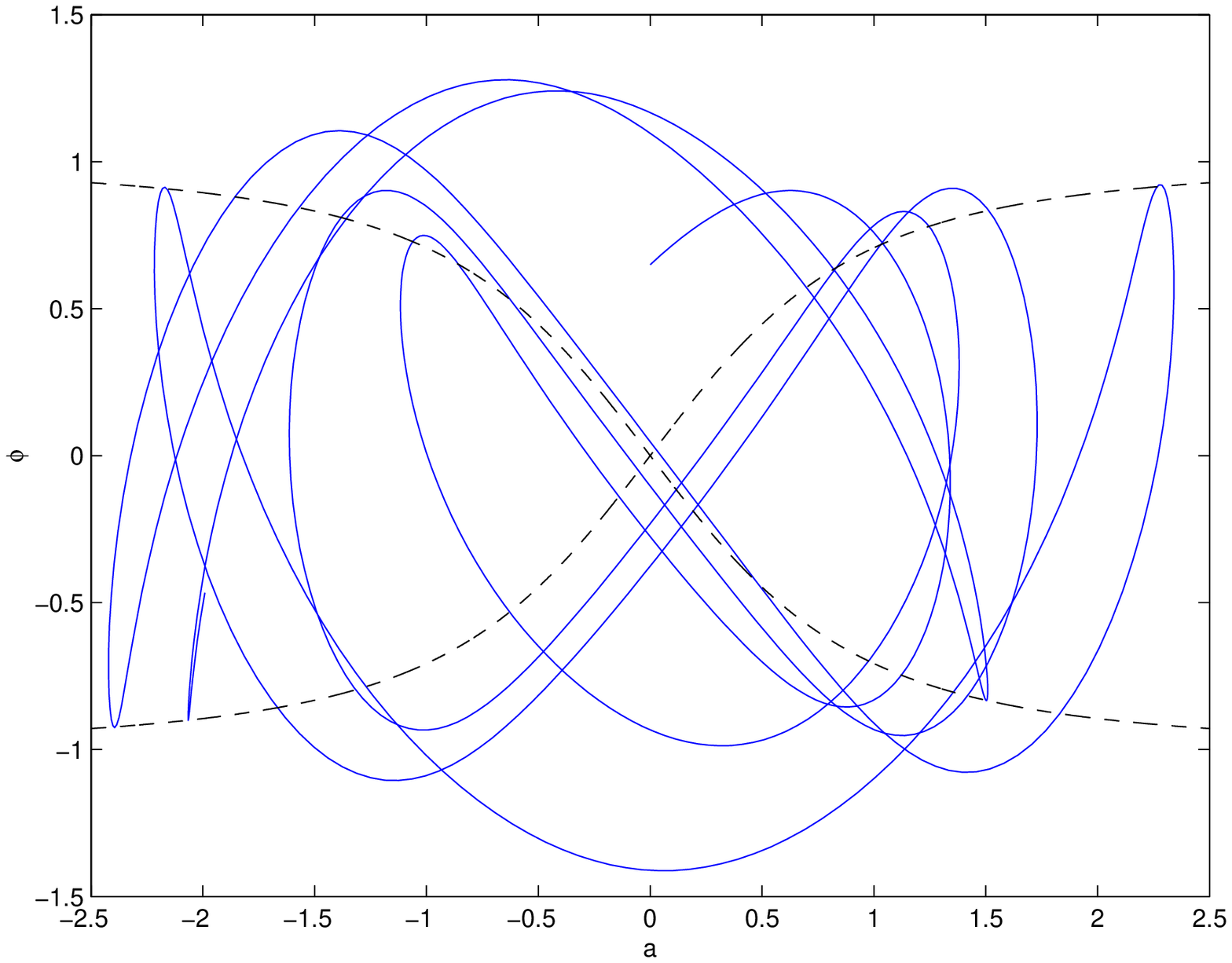}
\includegraphics[scale = 0.45]{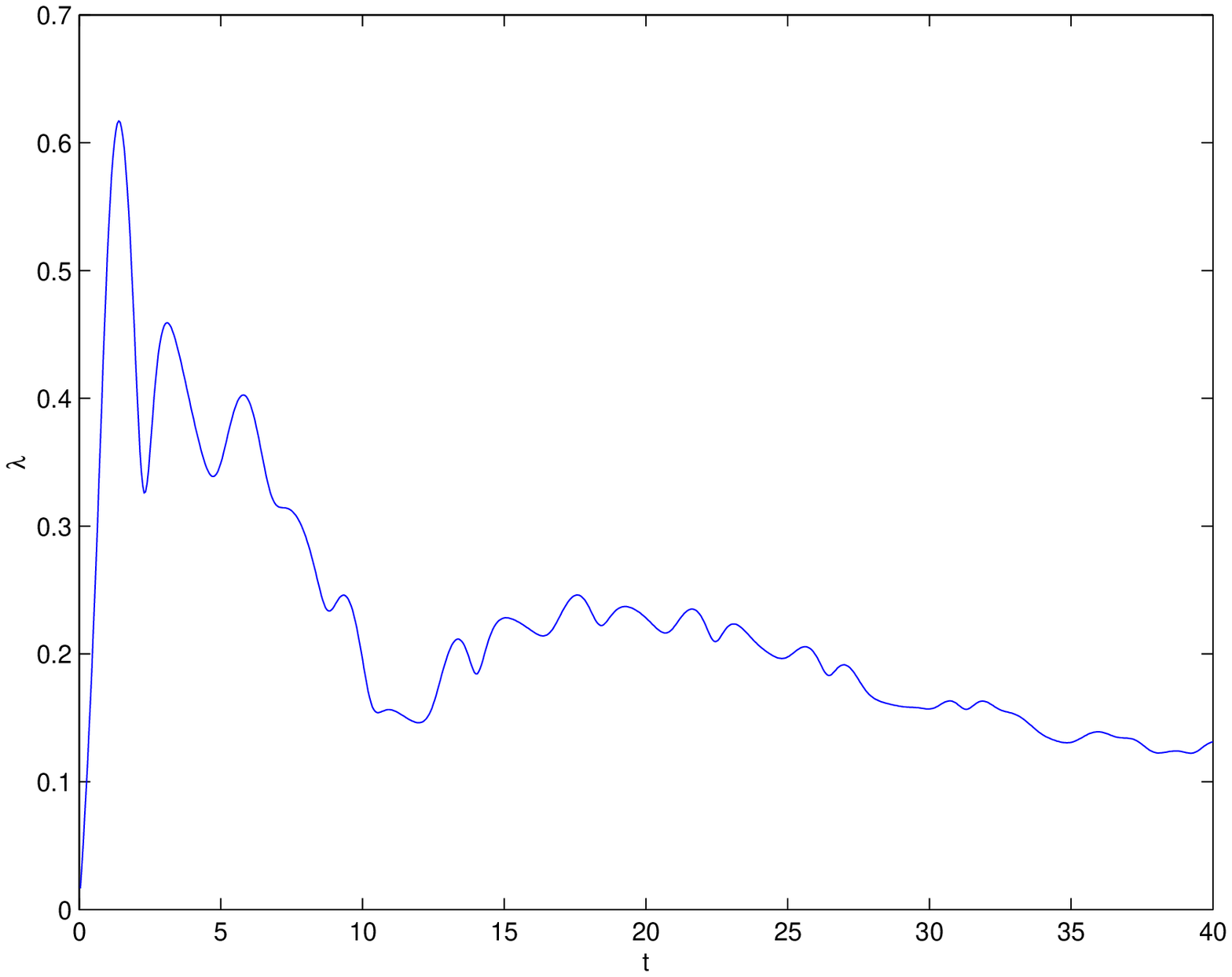}
\caption{The trajectory in the configuration space for initial conditions $a_{0}=0$, $\phi_{0}=\dot{\phi}_{0}=0.65$ ($\dot{a_{0}}$ calculated from Hamiltonian constraint) and Lyapunov principal exponent.}
\label{fig:4}
\end{center}
\end{figure}

Fig.~\ref{fig:2},\ref{fig:3},\ref{fig:4} shows three trajectories in the configuration space for three different initial conditions together with Lyapunov exponents in time, calculated in the standard way. Note that in all cases Lyapunov principal exponent goes to zero as $\eta \to \infty$ which corresponds to the singularity at the conformal time. This suggests that the system under consideration has no property of sensitive dependence on initial conditions which is characteristic rather for integrable systems. Another property which distinguishes the system from other systems of classical mechanics with conception of absolute time is that they do not posses property of topological transitivity which is crucial point for understanding of classical chaos conception \cite{Wiggins:1990}.

\begin{figure}
\begin{center}
\includegraphics[scale = 1]{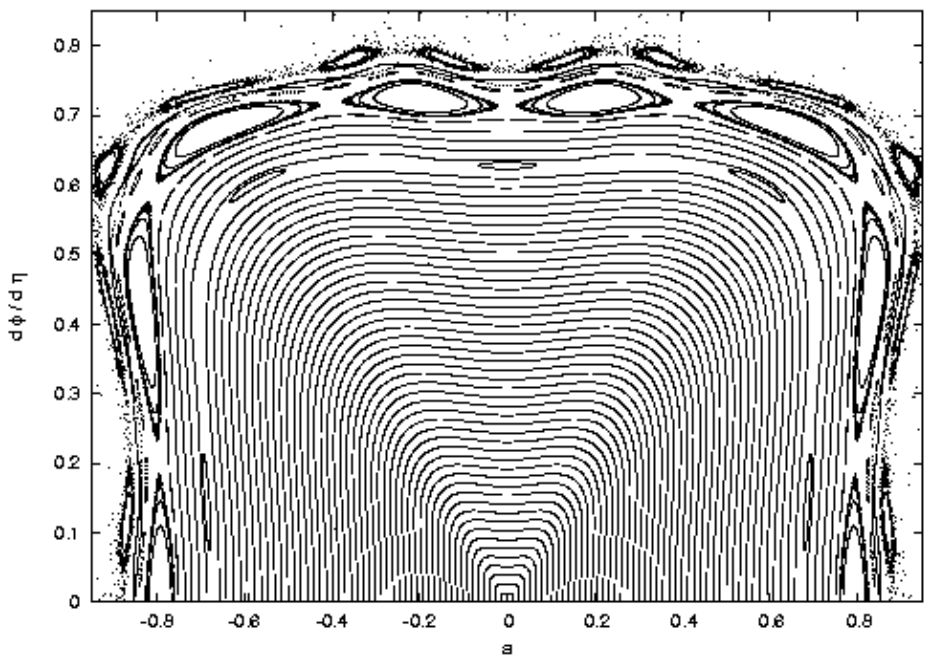}
\caption{Poincar{\'e} section surface $V(a,\phi)=0$ and $\dot{a} < 0$, $\dot{\phi} > 0$.}
\label{fig:5}
\end{center}
\end{figure}

It is useful to choose degeneration line $V(a,\phi)=0$ as a Poincar{\'e} surface. Let us concentrate on the simplest case of $k=+1$ and $\Lambda=\lambda=0$. Then $\phi=\pm a/\sqrt{1+a^{2}}$ is an algebraic equation of the boundary set. On Fig. \ref{fig:5} one can see Poincar{\'e} section on the surface $V(a,\phi)=0$ on the plane $(a,\dot{\phi})$ for different initial conditions.

\begin{figure}
\begin{center}
\includegraphics[scale = 1]{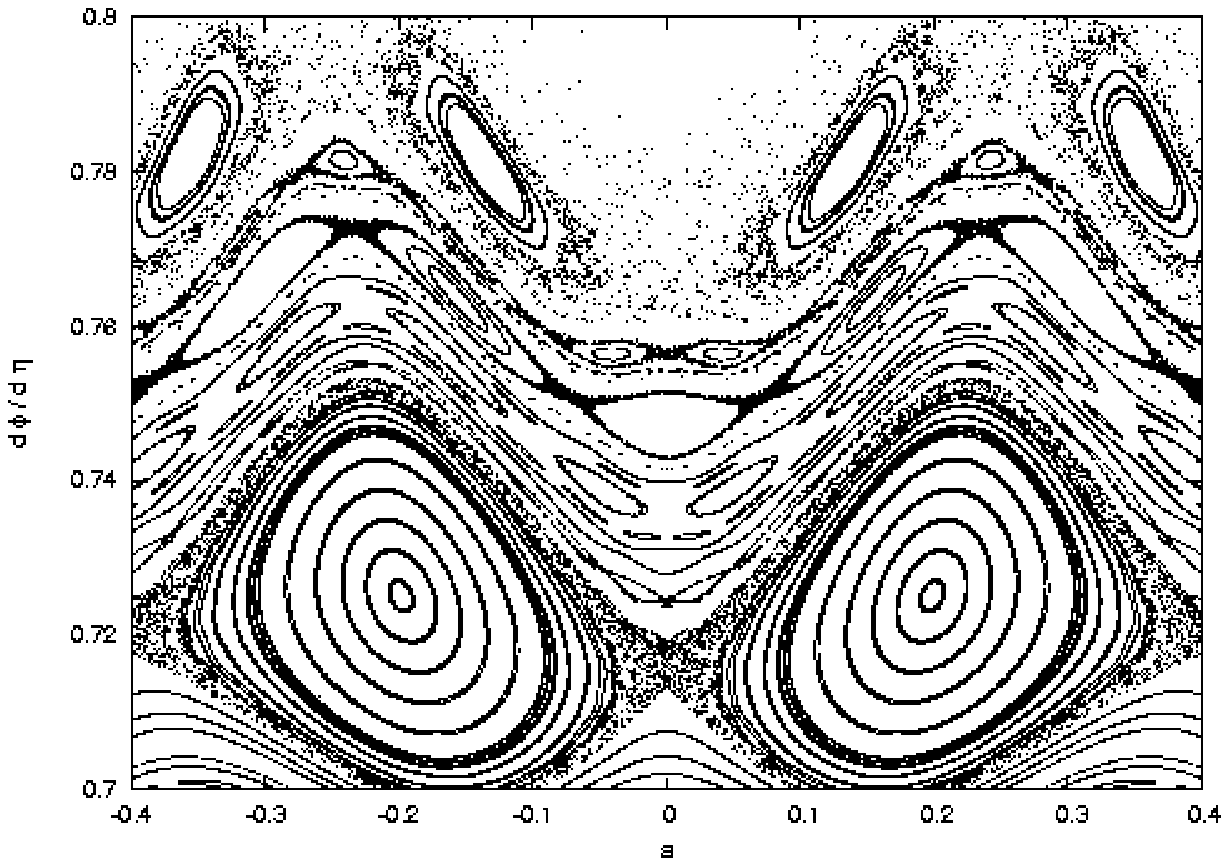}
\includegraphics[scale = 1]{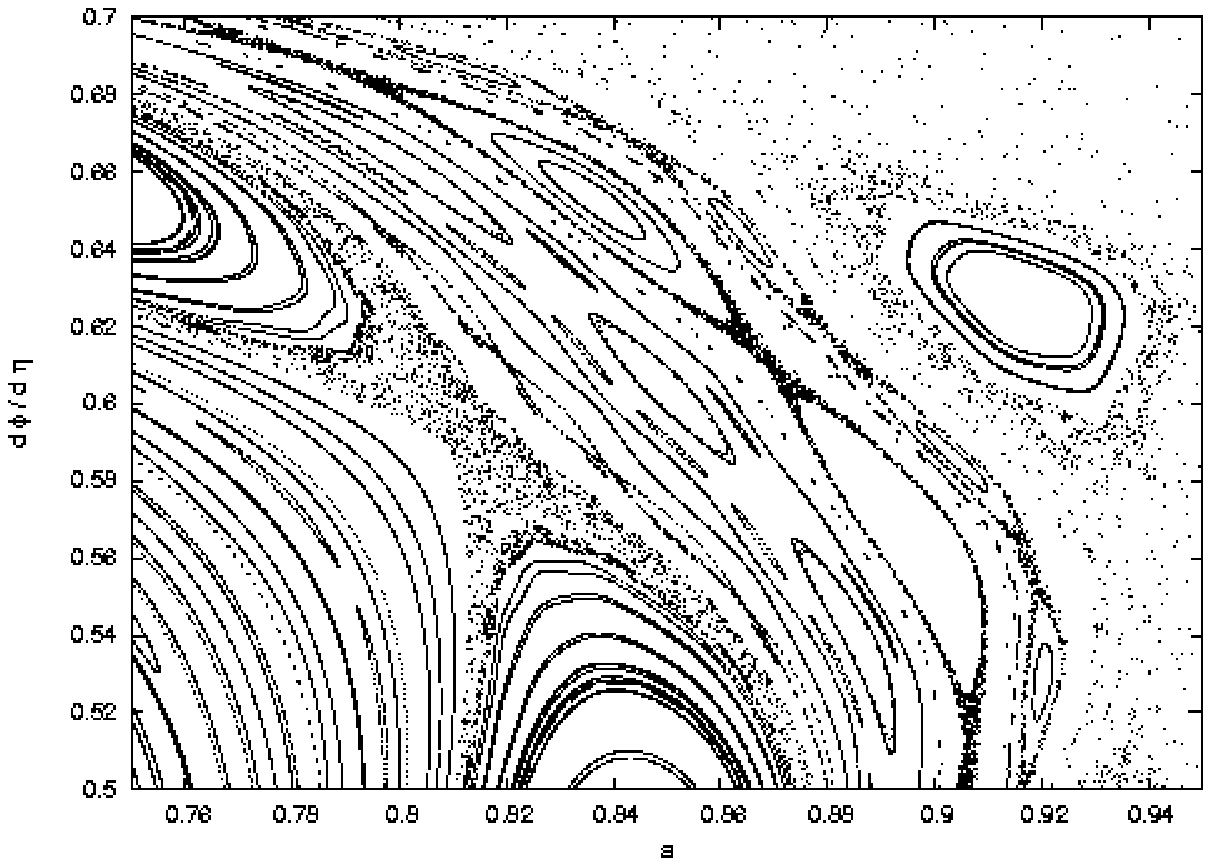}
\caption{Details of Poincar{\'e} section from Fig. \ref{fig:5}.}
\label{fig:7}
\end{center}
\end{figure}

While in Fig. \ref{fig:5} we find most trajectories as regular, there are, of course, chaotic trajectories. Some details of their concentrations in neighborhood of saddle points and forming stochastic layers is illustrated on Fig. \ref{fig:7}.

\subsection{Symbolic dynamics in detection of dynamical complexity}

Hadamard \cite{Hadamard:1898} was the first to use methods of trajectories coding in investigations of geodesics on compact space with negative curvature, which belongs now to the field of symbolic dynamics. The significance of this method in the context of closed cosmology with a scalar field was pointed out by Kamenshchik \textit{et~al.} \cite{Kamenshchik:1999}. The crucial feature of this model is the existence of points of maximal expansion ($\dot{a}=0$, $\ddot{a}<0$) and sometimes points of minimal contraction ($\dot{a}=0$, $\ddot{a}>0$) or ``bounces''. Then it is possible to classify all trajectories using localization of their points of maximal expansion and calculate the topological entropy which measures the growth of the number of periodical orbits as their period increases. Hence one can quantify the length of orbit by the number of symbols $A$ -- bounce of trajectory ($\dot{a}=0$, $\ddot{a}>0$), $B$ -- crossing the line $\phi=0$. Note that $a \ge 0$ for physical reasons the extension of trajectories to $a<0$ domain means the extending the solutions beyond the big crunch which is only mathematically admissible \cite{Motter:2002}.

For the model under consideration, two different coding procedures were used. In the first method we count all the intersections of trajectories with the axis $\{a=0\}$. For $\phi>0$ we put symbol $(1)$ while in opposite case if $\phi<0$ we put symbol $(0)$. Another method of codding trajectories rests on the analysis of intersections points with the boundary set $\partial\mathcal{D}$ defined as the surface $V(a,\phi)=0$ in the configuration space $(\phi,a)$. One can quantify the length of the orbit by two symbols : $(1)$ -- if $\phi>0$ at intersection point, and $(0)$ in the opposite case if $\phi<0$. In both approaches mentioned before the trajectories are represented by a sequence of zeros and ones. The next step in our analysis is the division of all coding trajectories into blocks in the simplest way. The blocks consist of to letters $(0 0)$, $(0 1)$, $(1 0)$ and $(1 1)$ and blocks $(0 1)$ and $(1 0)$ are treated as a identical. Then, after counting different blocks one can calculate the Shannon informative entropy following the rule
\begin{equation}
H_{S} = \sum_{i=1}^{r} p_{i} \ln{\frac{1}{p_{i}}},
\label{eq:20}
\end{equation}
where $p_{i}$ $(i=1,\dots,r)$ is the probability that a given block will appear in the trajectory coding.

\begin{figure}
\begin{center}
\includegraphics[scale=0.75]{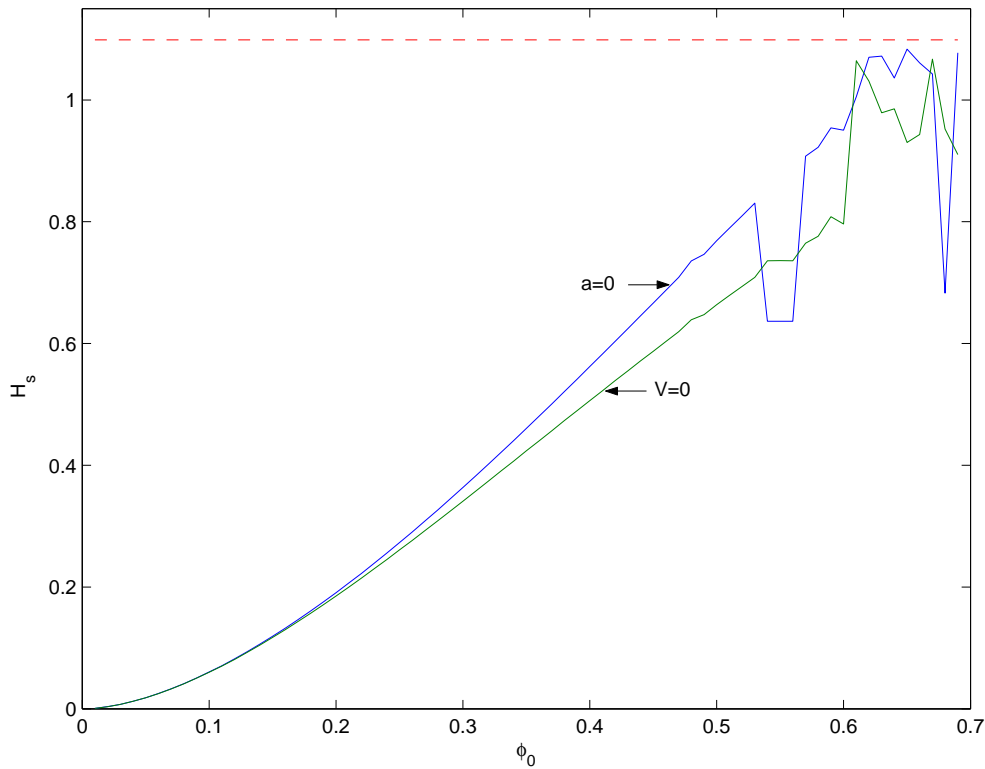}
\caption{Shannon informative entropy as a function of $\phi_{0}$ for two different codding methods. The horizontal line $H_{S}=\ln{3}$ denotes the limit for purely random process.}
\label{fig:8}
\end{center}
\end{figure}

The informative Shannon entropy characterizes the uncertainty degree of appearance of a given result. Following definition (\ref{eq:20}) $H_{S}$ is a number which belongs to the interval $[0, \ln{r}]$. If for all $i$ $p_{i}$ assumes the same value equal $1/r$ then from definition (\ref{eq:20}) we obtain $H_{S}=\ln{r}$ which is the maximal value of $H_{S}$. From Fig. \ref{fig:8} one can observe that $H_{S}$ is a growing function of initial conditions $\phi_{0}$ to the limit which corresponds to purely chaotic behavior.

\subsection{Distribution of intersection points on the boundary set $\partial\mathcal{D}$ as a measure of complexity of dynamics}

\begin{figure}
\begin{center}
\includegraphics[scale=0.45]{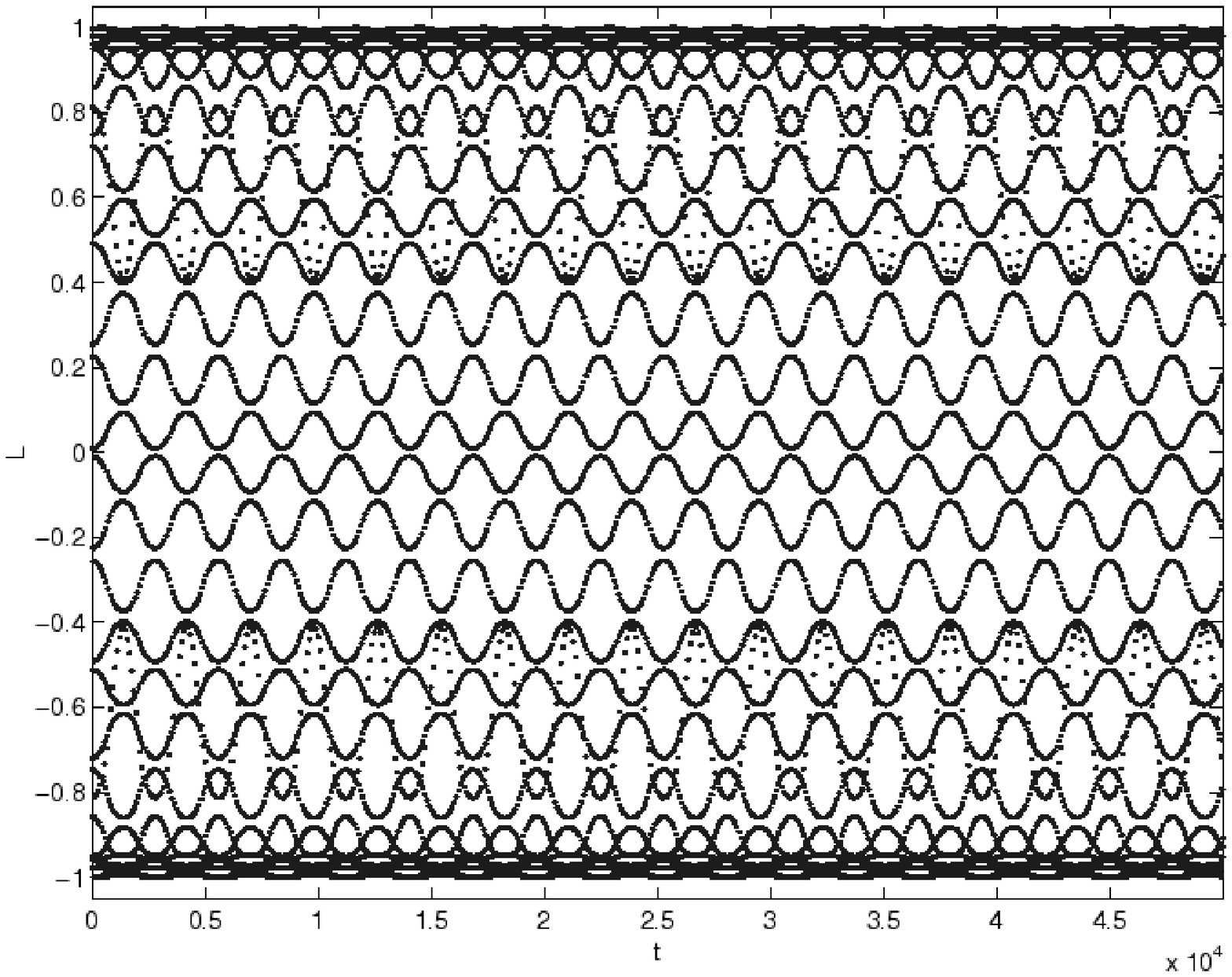}
\includegraphics[scale=0.45]{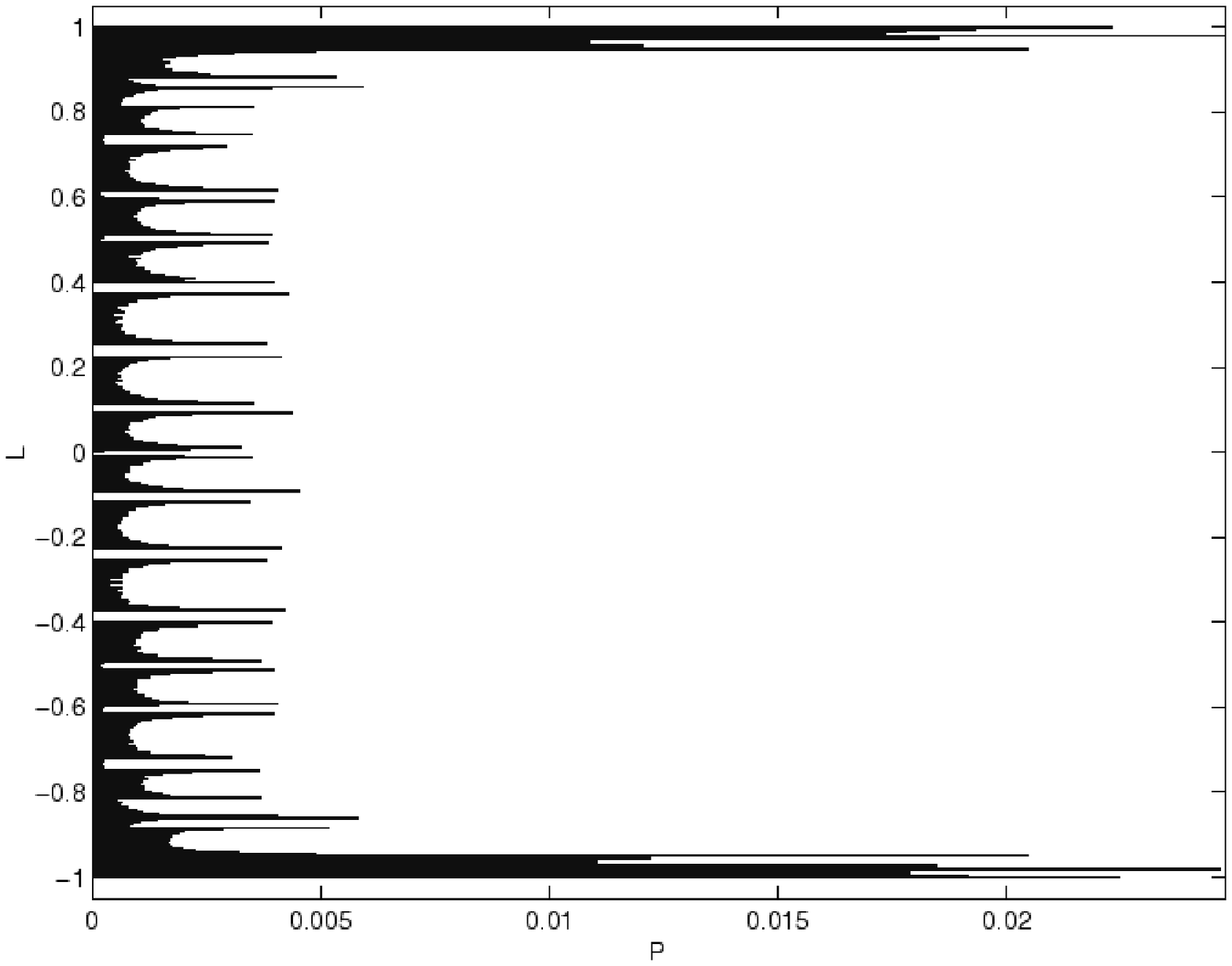}

\includegraphics[scale=0.45]{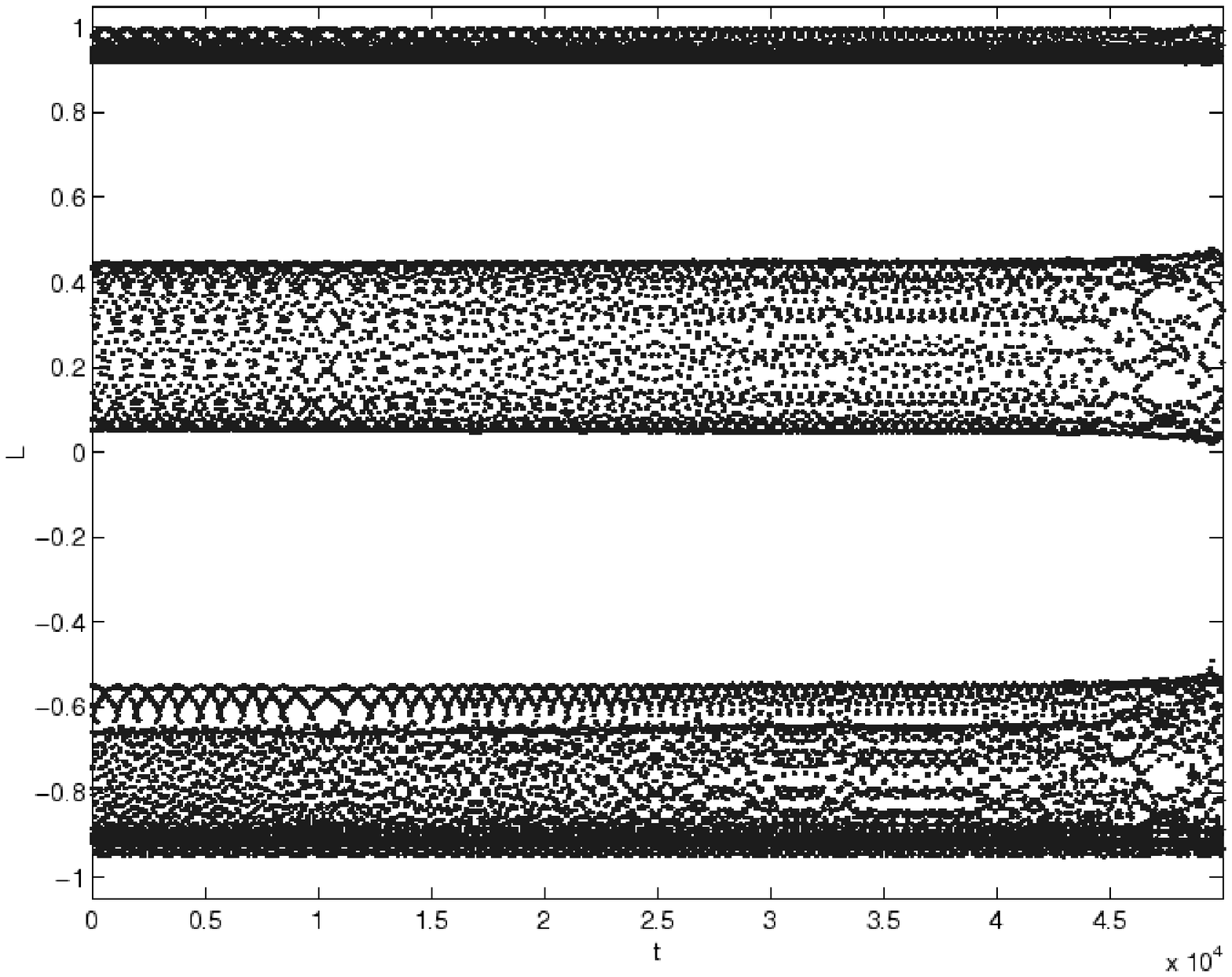}
\includegraphics[scale=0.45]{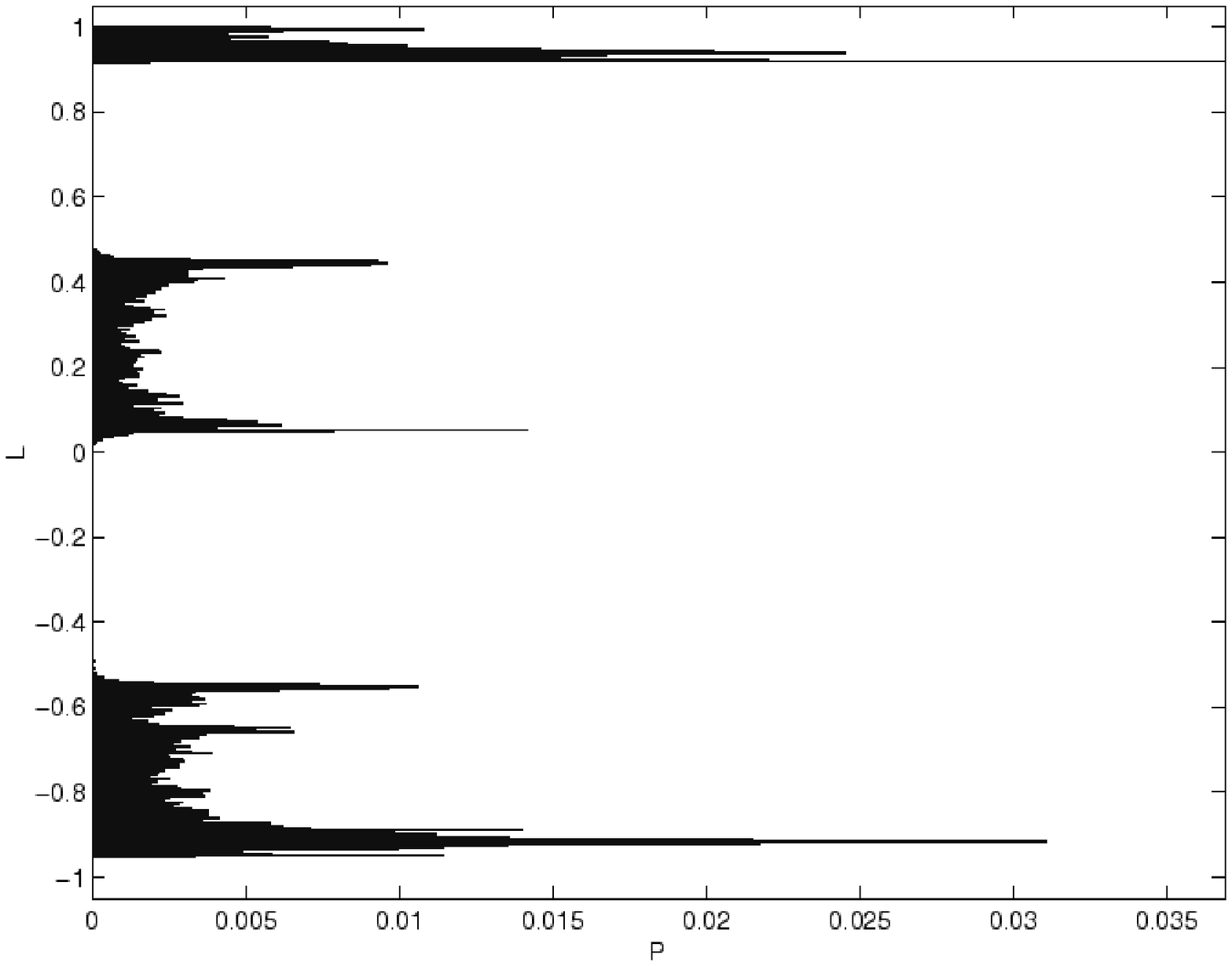}
\caption{Intersection points of the boundary set $\partial\mathcal{D}$. Points of intersection of trajectory with set $\partial\mathcal{D}$ as a function of time (left) and probability of finding a particle-universe on a given subset of the set $\partial\mathcal{D}$ during evolution (right).}
\label{fig:9}
\end{center}
\end{figure}

\begin{figure}
\begin{center}
\includegraphics[scale=0.45]{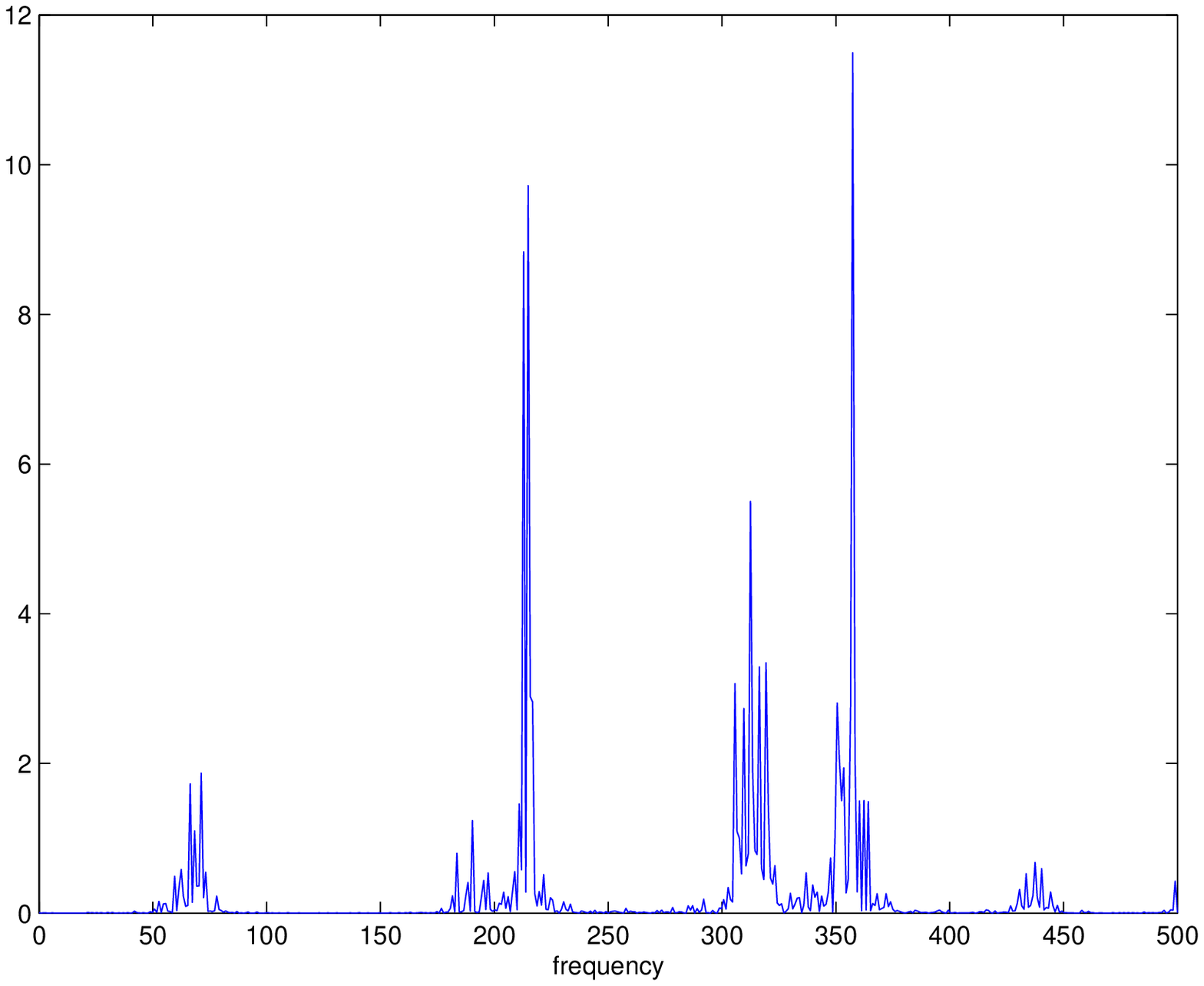}
\includegraphics[scale=0.45]{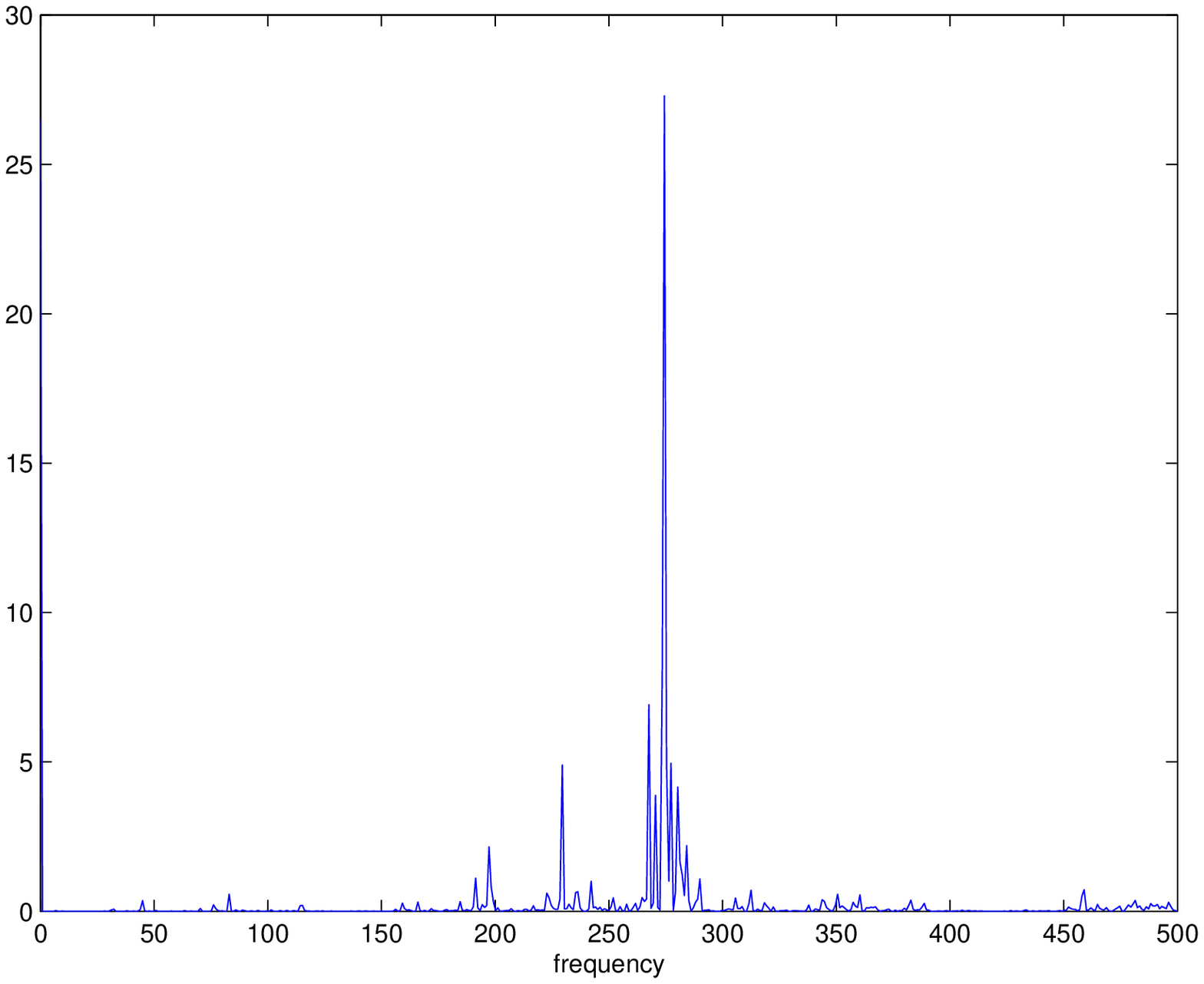}
\caption{Fourier analysis of distributions form Fig. \ref{fig:9}. Existence of weak noise in the power spectrum can indicate chaotic distribution of intersection points.}
\label{fig:10}
\end{center}
\end{figure}

It is interesting to check how the intersection points are distributed on the boundary line described by function $\phi=a/\sqrt{1+a^{2}}$. Let $L$ be the length along that line calculated form the origin to the intersection point and $N$ denote number of such intersections in the interval $L \pm \Delta L$. Of course $N$ can be normalized to $P$ and treated as a probability of finding a fictitious particle moving along a geodesic at a given point of $\partial\mathcal{D}$. The probability $P$ as a function of $L$ (normalized to unity) is shown in Fig. \ref{fig:9}. For deeper analysis of the distribution of the points Fourier analysis was performed and the results of such analysis are shown in Fig. \ref{fig:10}. The existence of weak noise in the power spectrum can indicate chaotic distribution of the intersection points.

\subsection{The existence of unstable periodic orbits as an indicator of complexity of dynamics}

It would by useful for classification of the periodic orbits to consider some reflection symmetries of the system. Of course, the system possesses reflection symmetry $t \to -t$ and $q^{i} \to -q^{i}$ $(q^{i}=a,\phi)$. Therefore it is sufficient to investigate motion in one quarter, say $a>0$ and $\phi>0$, to reconstruct the motion in the remaining quarters.

\begin{figure}[t]
\begin{center}
\includegraphics[scale=0.45]{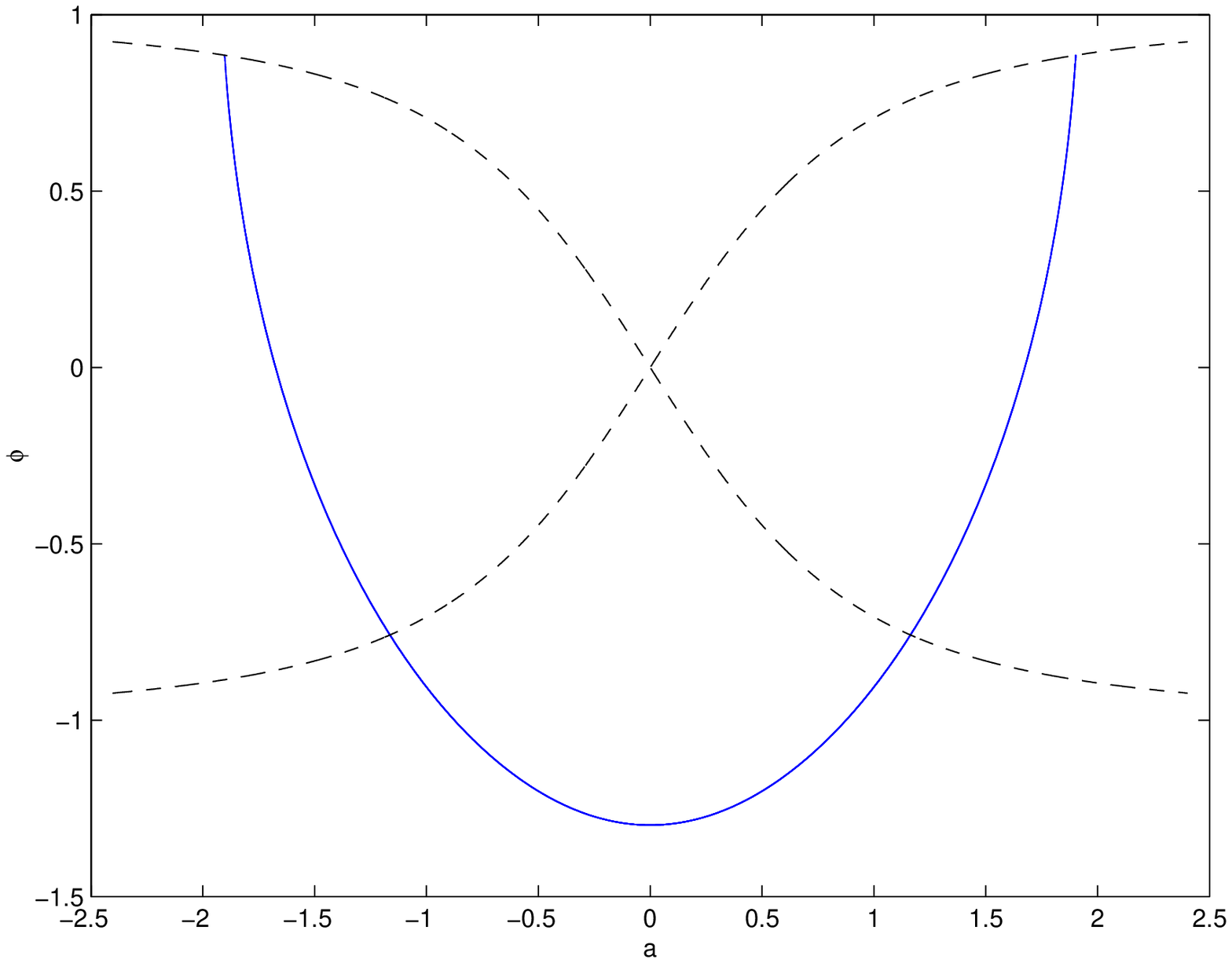}
\includegraphics[scale=0.45]{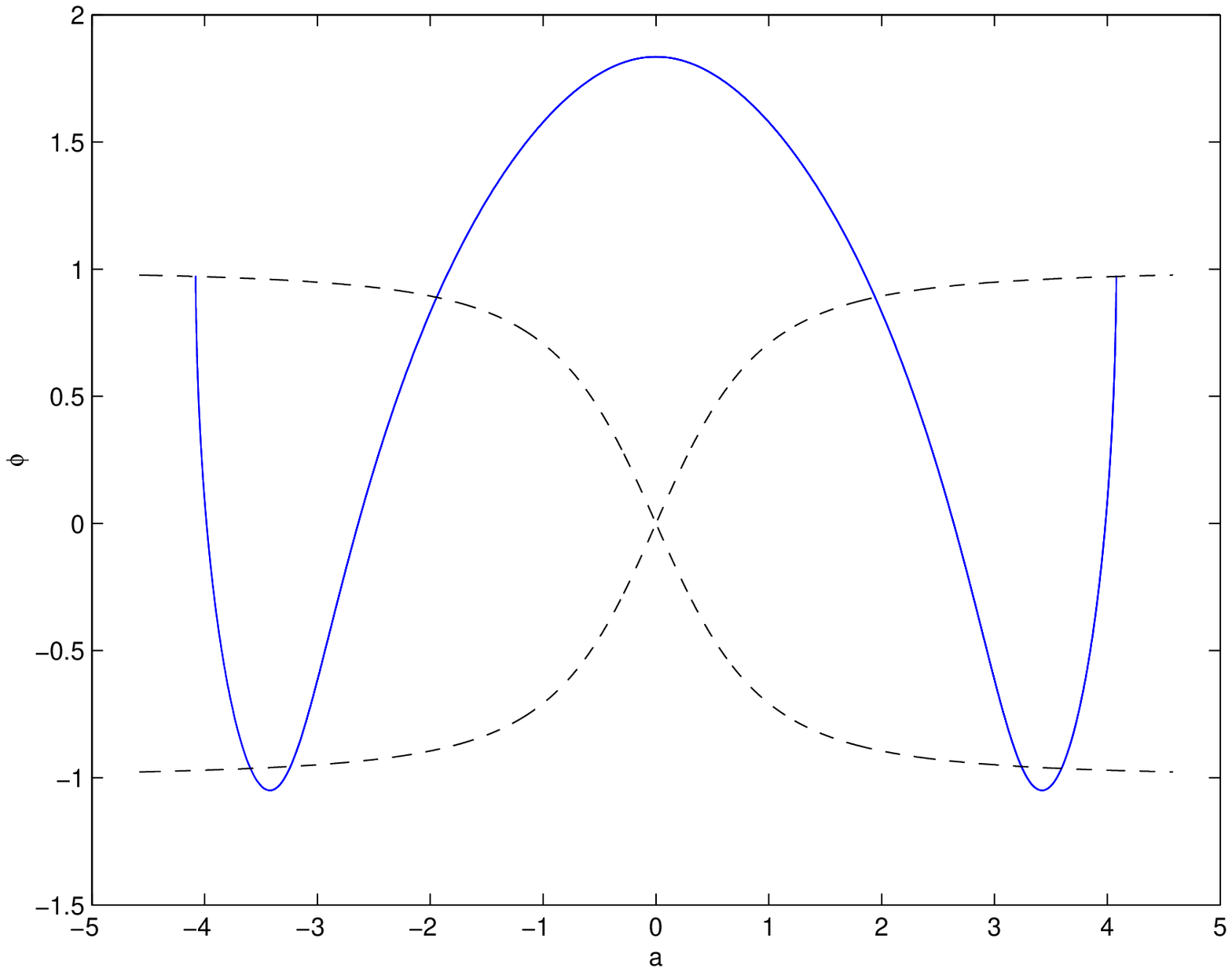}
\caption{Periodic trajectories of type `I a' in the configuration space.}
\label{fig:11}
\end{center}
\end{figure}

\begin{figure}[t]
\begin{center}
\includegraphics[scale=0.45]{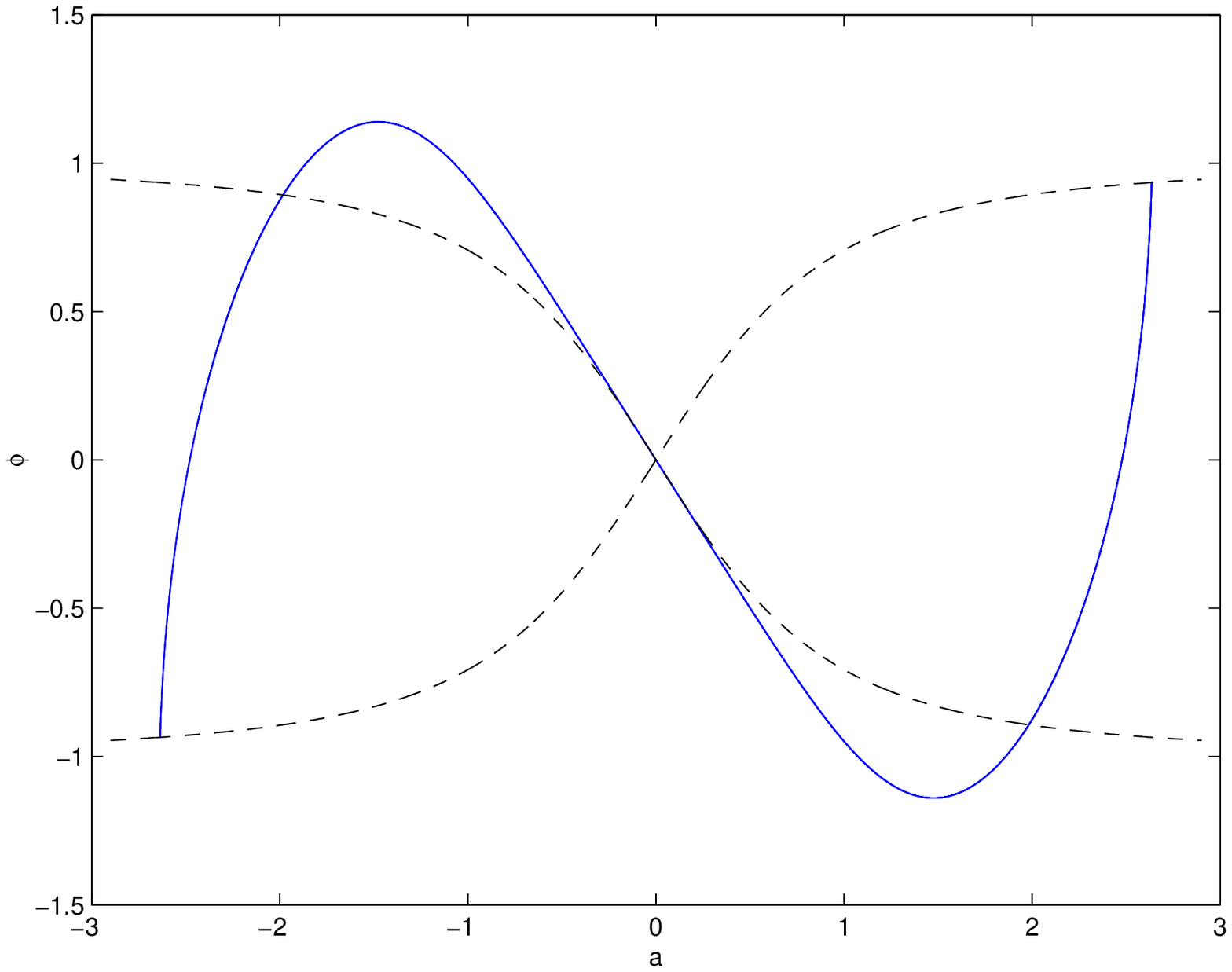}
\includegraphics[scale=0.45]{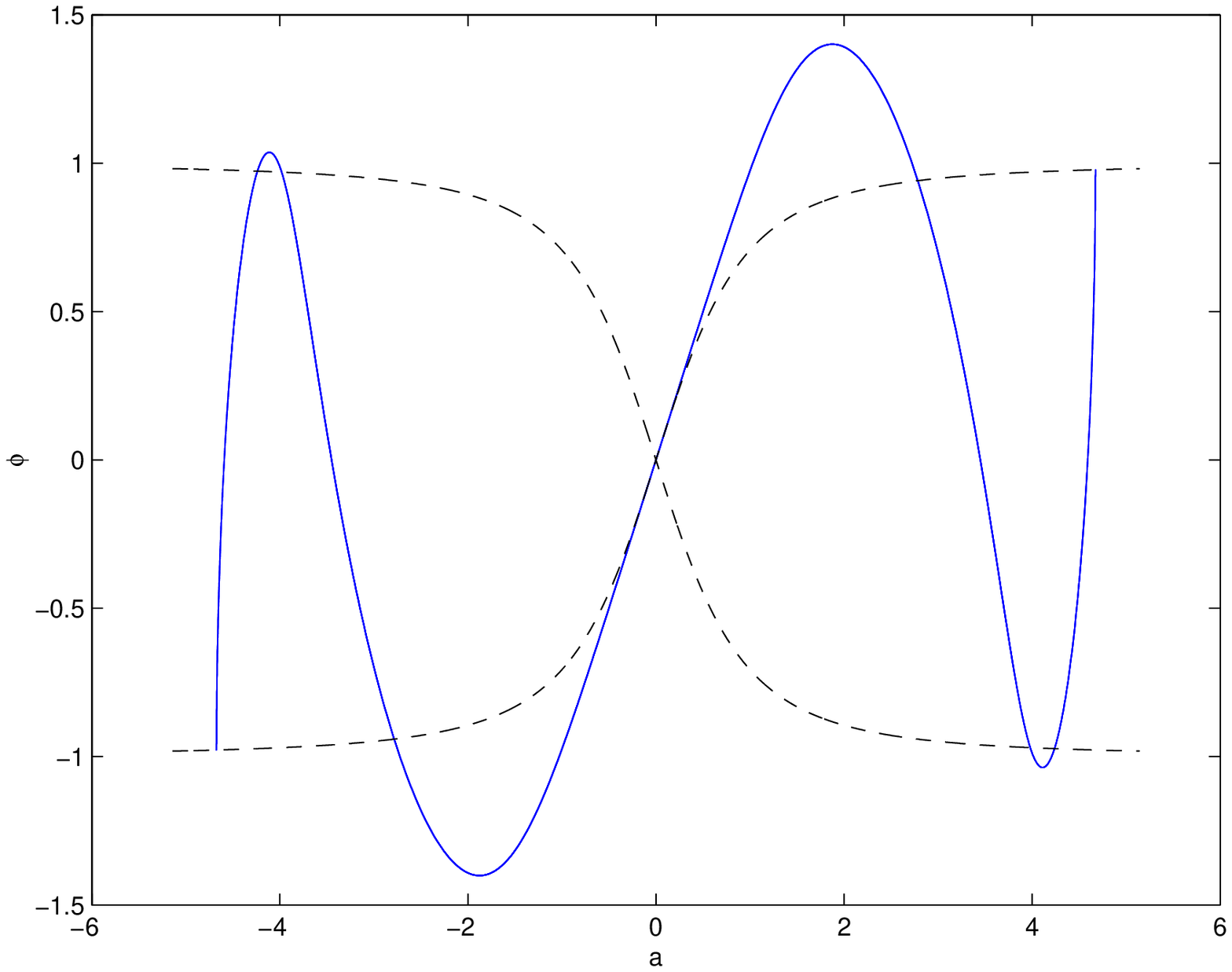}
\caption{Periodic trajectories of type `I b' in the configuration space.}
\label{fig:12}
\end{center}
\end{figure}

\begin{figure}[t]
\begin{center}
\includegraphics[scale=0.45]{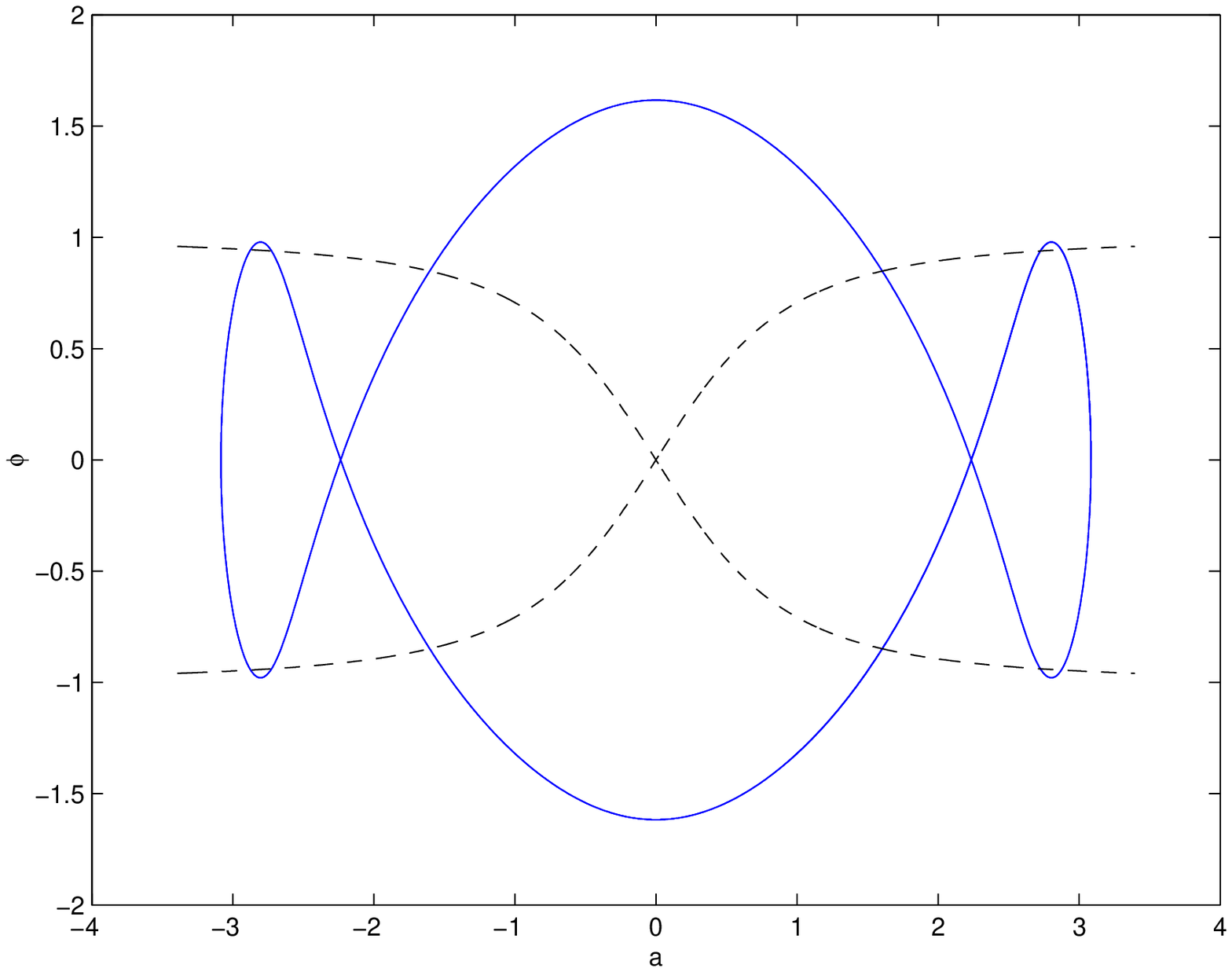}
\includegraphics[scale=0.45]{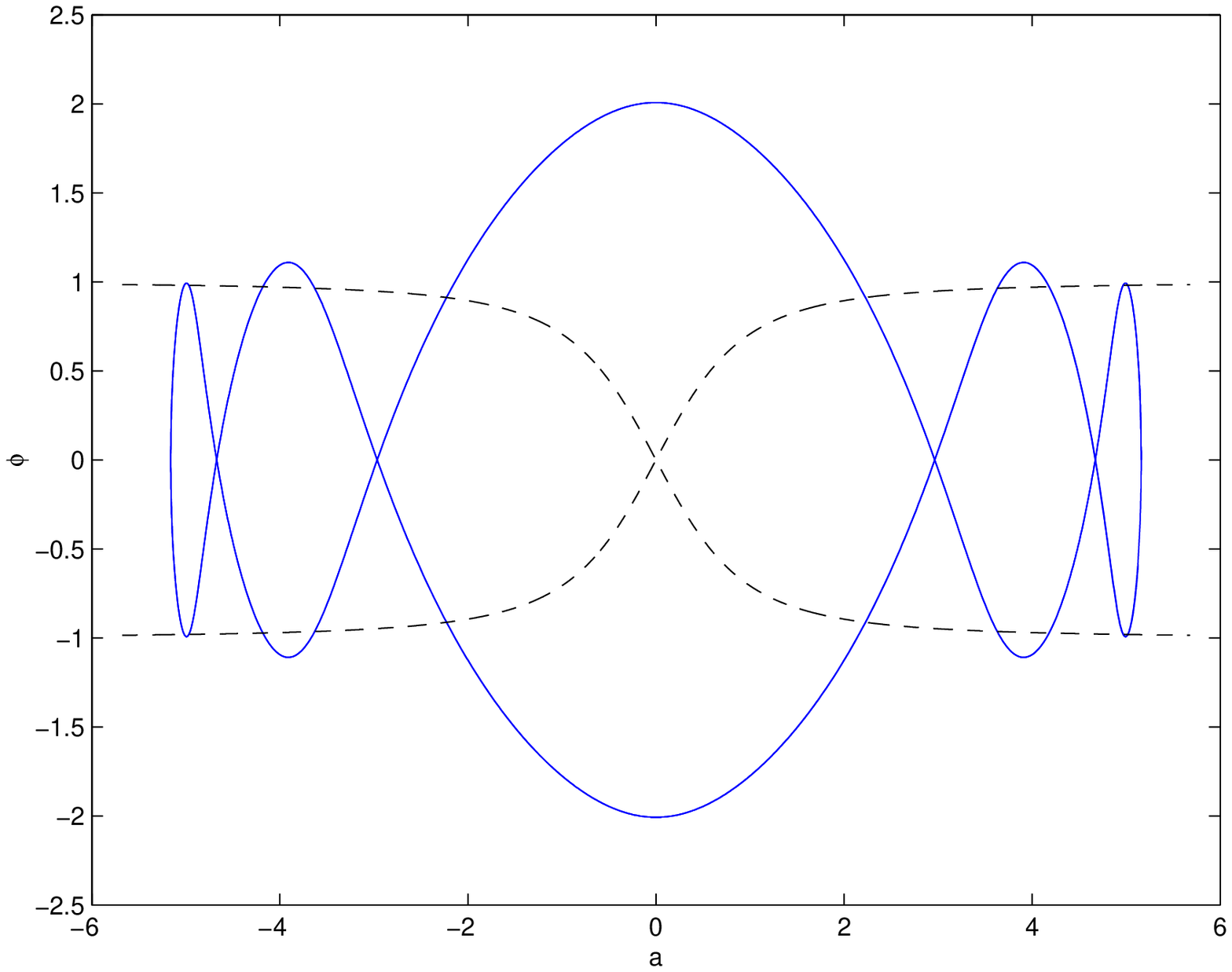}
\caption{Periodic trajectories of type `II a' in the configuration space.}
\label{fig:13}
\end{center}
\end{figure}

\begin{figure}[t]
\begin{center}
\includegraphics[scale=0.45]{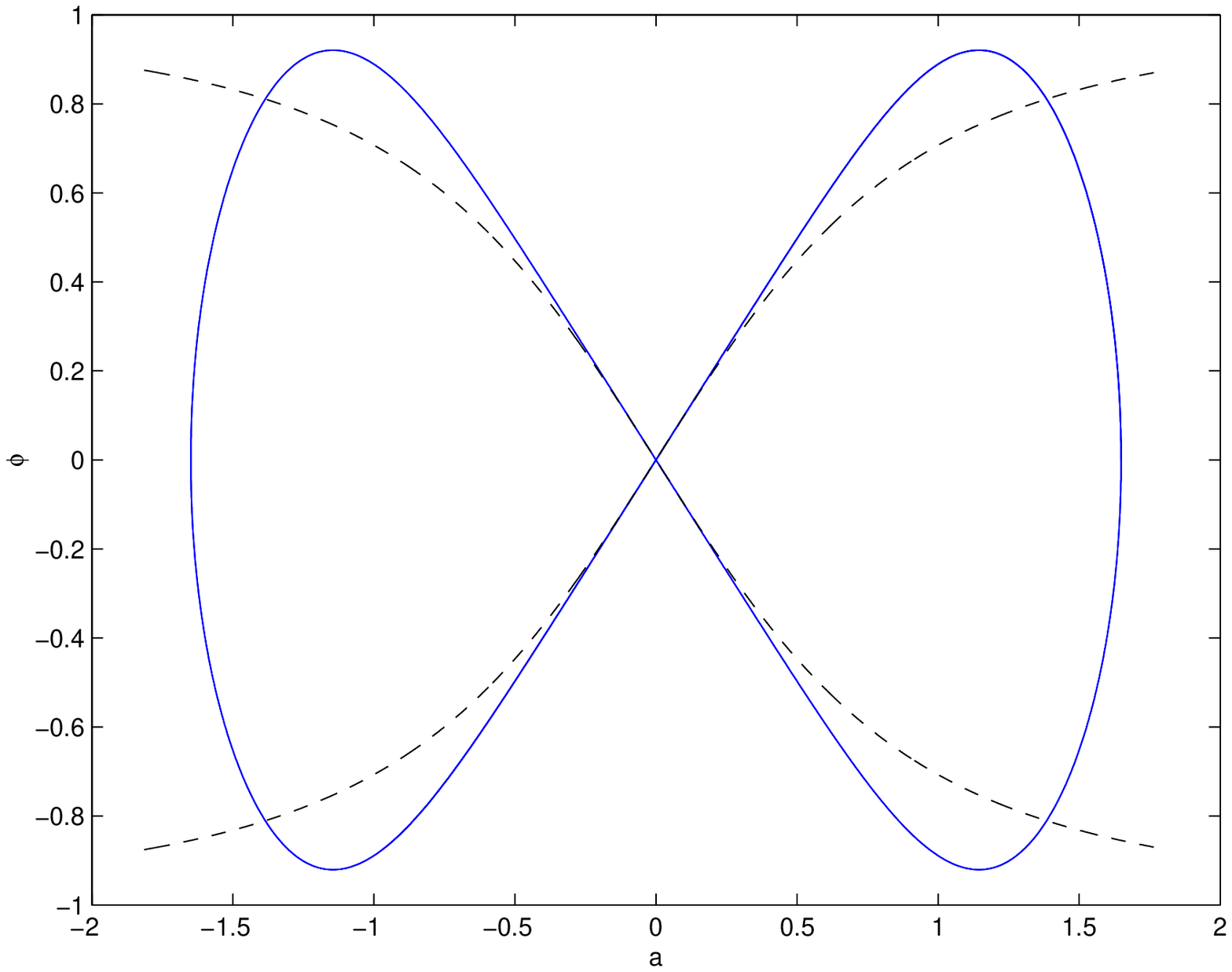}
\includegraphics[scale=0.45]{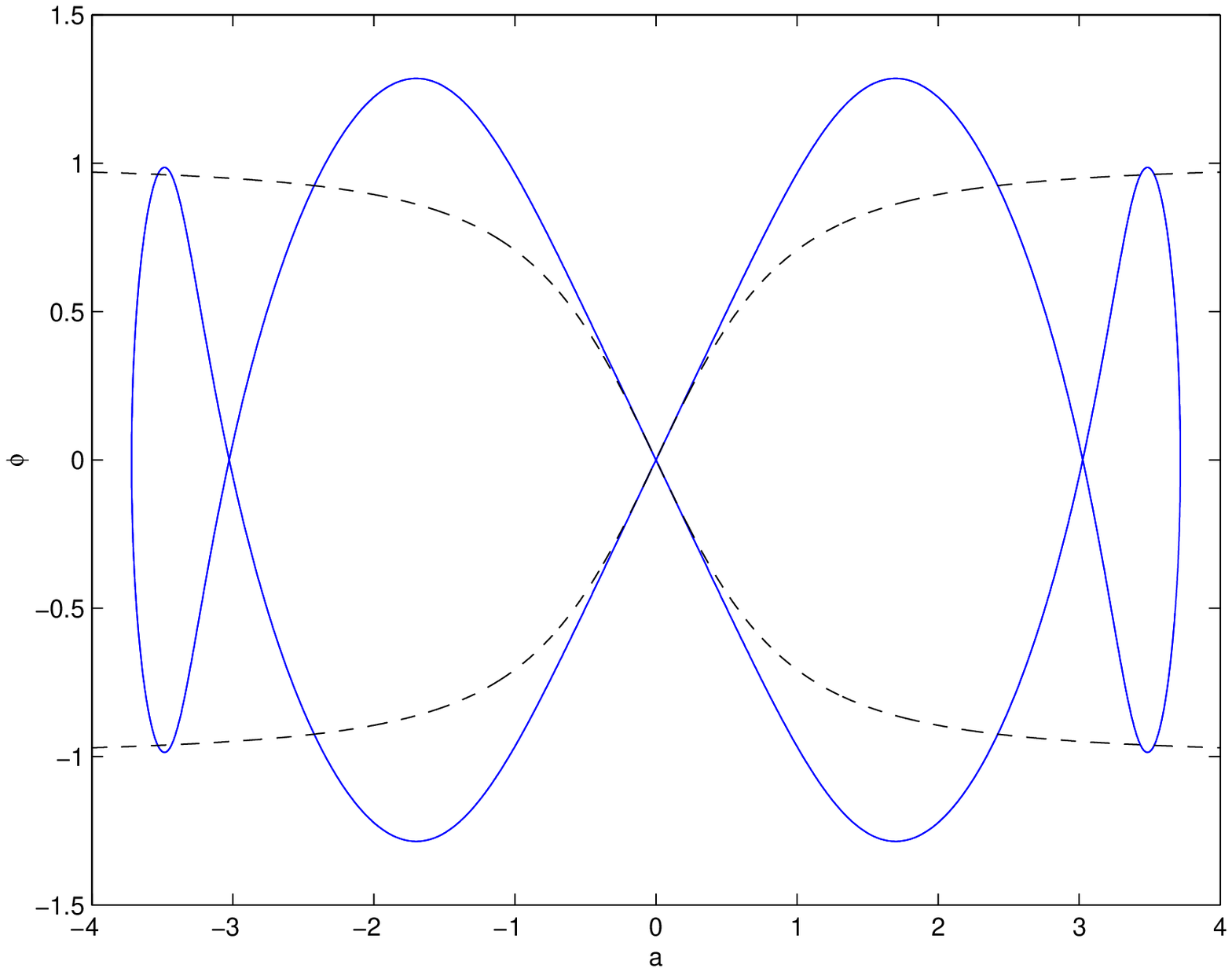}
\caption{Periodic trajectories of type `II b' in the configuration space.}
\label{fig:14}
\end{center}
\end{figure}

\begin{figure}[t]
\begin{center}
\includegraphics[scale=0.45]{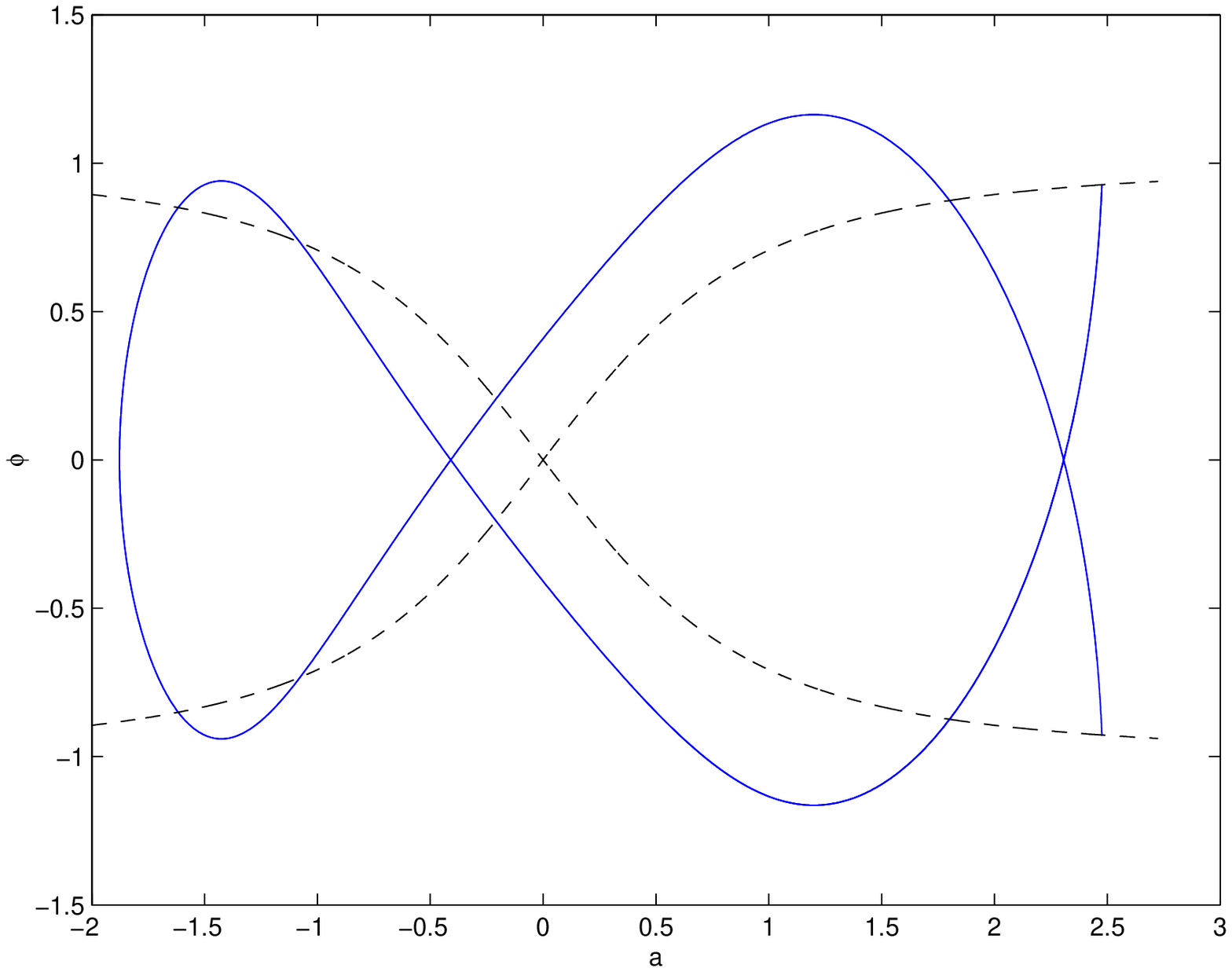}
\includegraphics[scale=0.45]{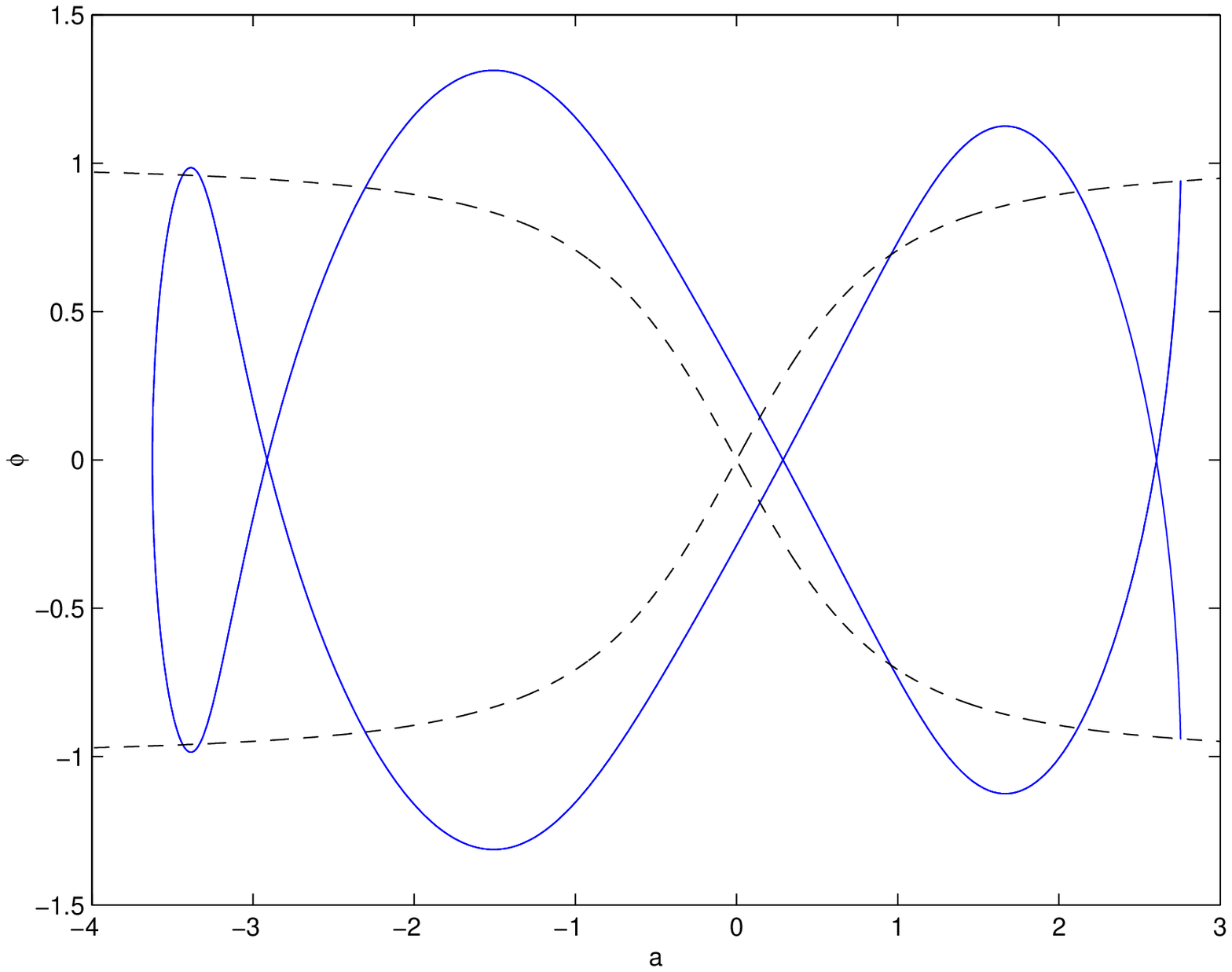}
\caption{Periodic trajectories of type `III' in the configuration space.}
\label{fig:15}
\end{center}
\end{figure}

\begin{figure}[t]
\begin{center}
\includegraphics[scale=0.45]{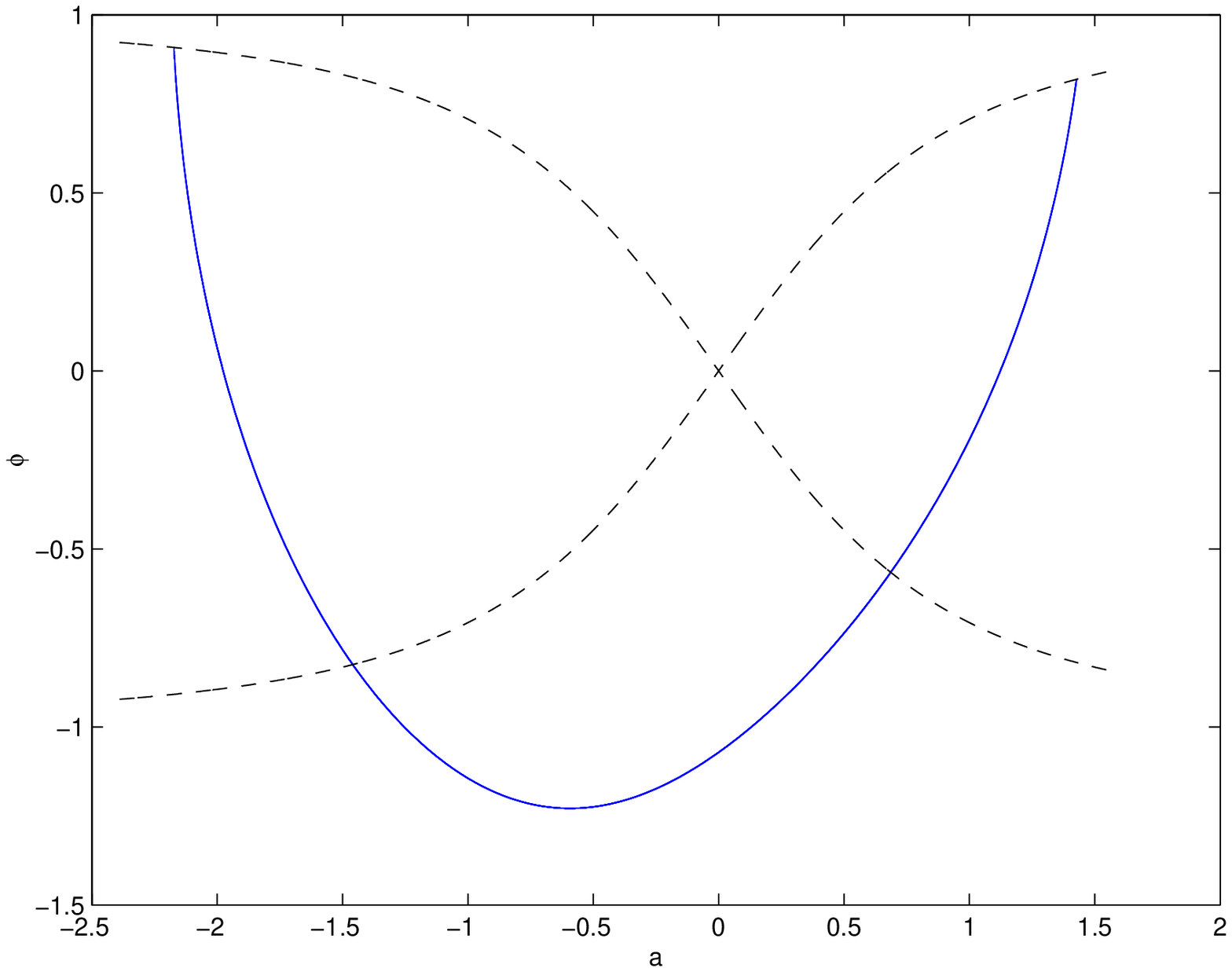}
\includegraphics[scale=0.45]{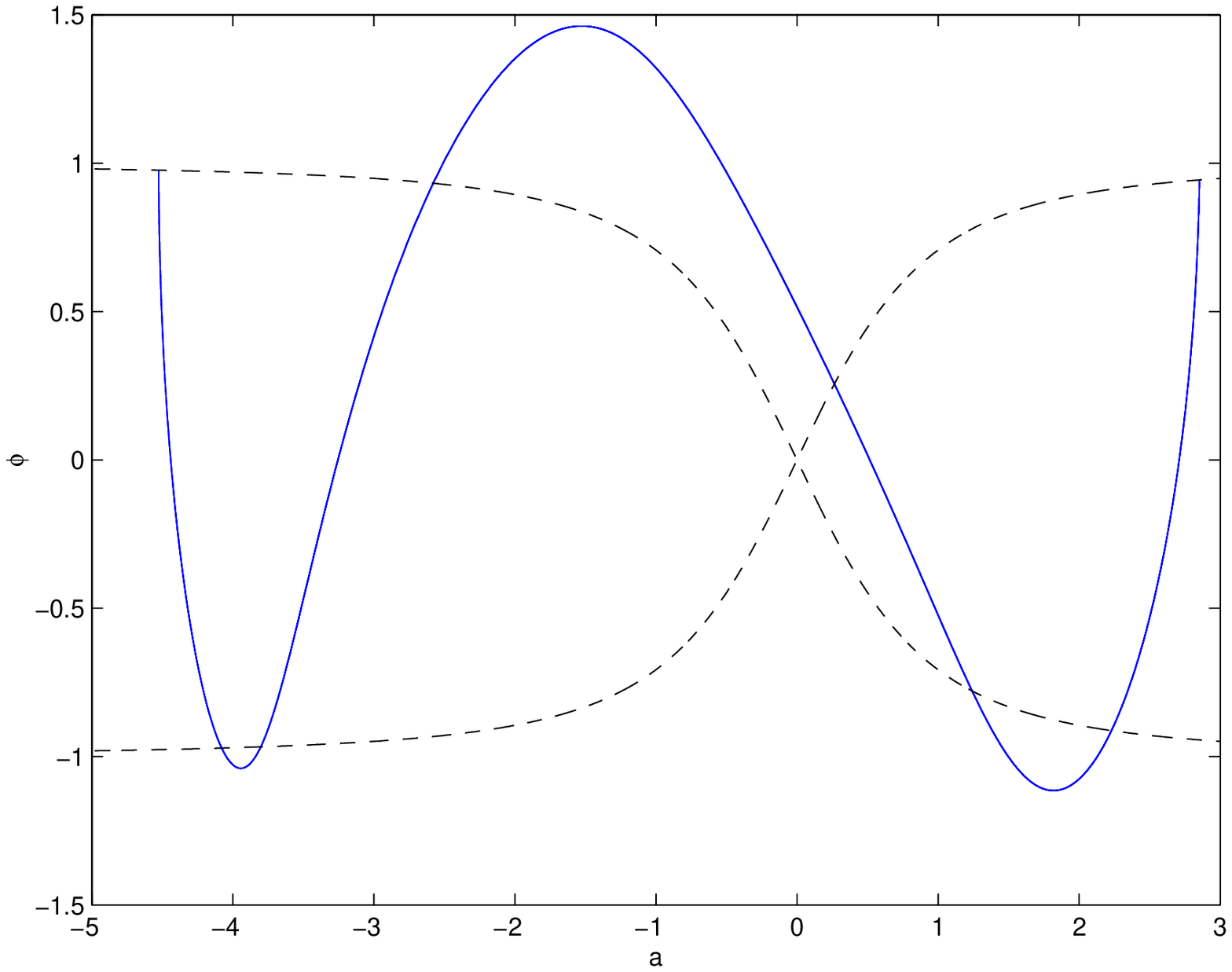}
\caption{Non-symmetrical periodic trajectories in the configuration space.}
\label{fig:16}
\end{center}
\end{figure}

\begin{figure}[t]
\begin{center}
\includegraphics[scale=0.45]{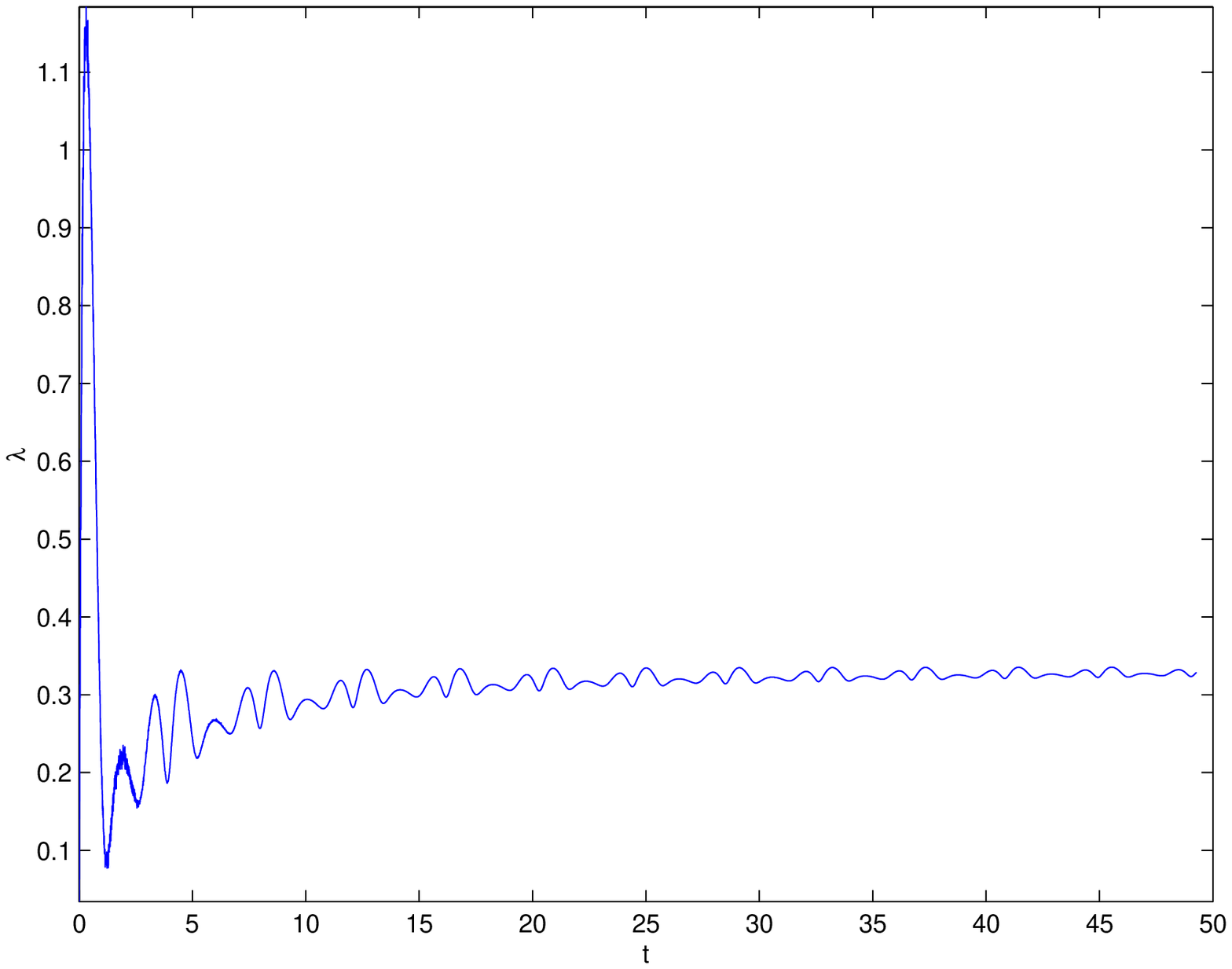}
\includegraphics[scale=0.45]{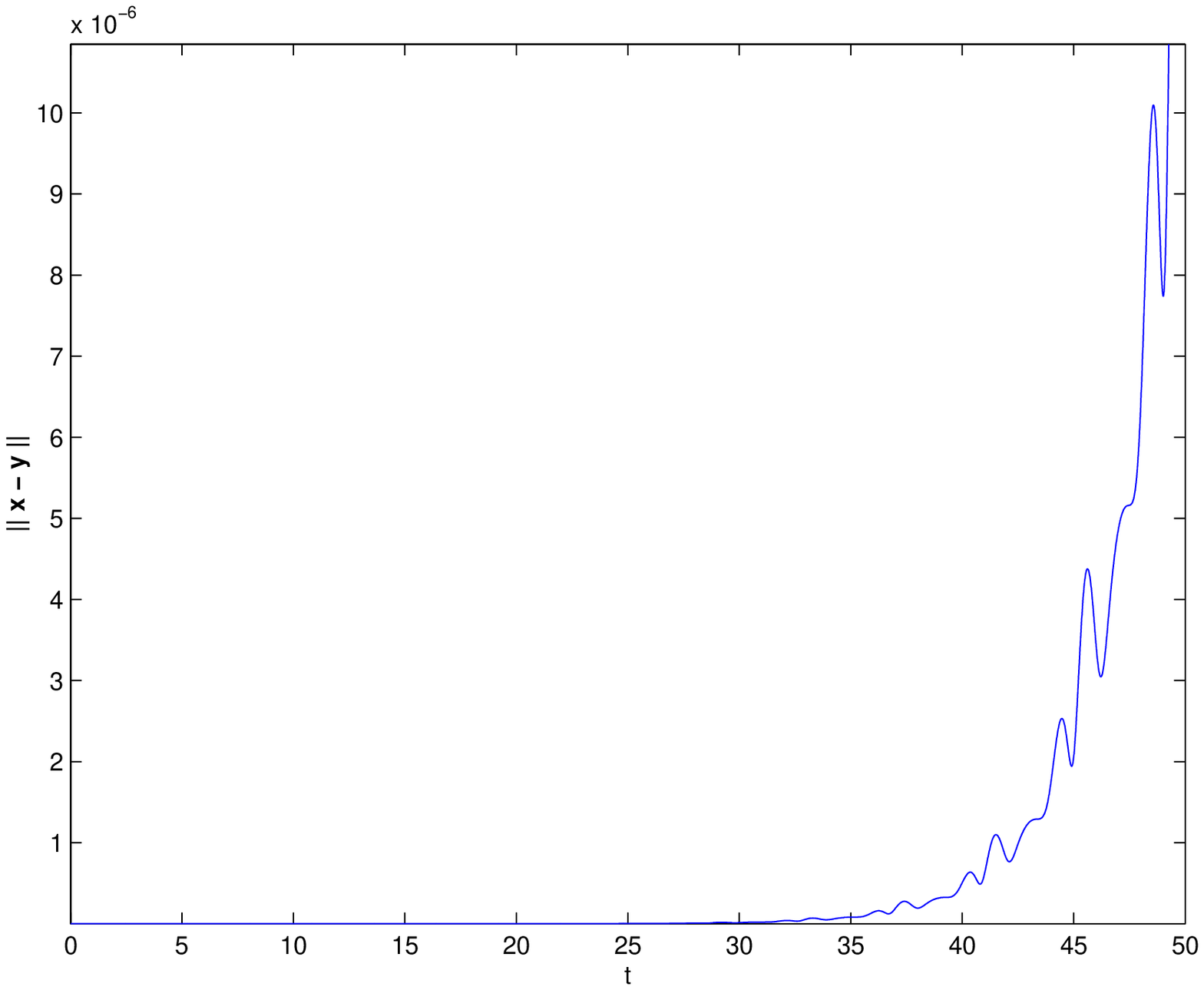}
\caption{Stability analysis for the first trajectory from Fig. \ref{fig:11}. Principal Lyapunov exponent (left) and the distance of separation vector (right) for initial conditions' separation $\Delta a = 10^{-12}$.}
\label{fig:17}
\end{center}
\end{figure}

Due to the symmetries mentioned before the simplest periodic orbits in the configuration space can be grouped in the following five classes
\begin{itemize}
\item{trajectories of type `I a' starting form the boundary line $V(a_{0},\phi_{0})=0$ with initial conditions $(\dot{a}_{0},\dot{\phi}_{0}) = (0,0)$ for which after a $1/4$ of period the inflection point at $\phi$ at $a=0$ is reached with $(\dot{a},\dot{\phi})=(-\phi,0)$ or $(+\phi,0)$; see Fig. \ref{fig:11}}
\item{trajectories of type 'I b' also starting from the boundary set and reaching the singular point $(a,\phi)=(0,0)$ for which $\dot{a}=\dot{\phi}$ or $\dot{a}=-\dot{\phi}$ after a $1/4$ of the full period; see Fig. \ref{fig:12}}
\item{trajectories of type 'II a' starting from a point of configuration space $(a_{0},0)$ with $(\dot{a}_{0},\dot{\phi}_{0})=(0,a_{0})$ and arriving at the inflection point of $\phi(a)$ diagram at $a=0$ $(\dot{a},\dot{\phi})=(-\phi,0)$ or $(+\phi,0)$ after a $1/4$ of full period; see Fig. \ref{fig:13}}
\item{trajectories of type 'II b' starting form the point $(a_{0},0)$, $(\dot{a}_{0},\dot{\phi}_{0})=(0,a_{0})$ and reaching the singular point $(a,\phi)=(0,0)$ and $\dot{a}=\dot{\phi}$ or $\dot{a}=-\dot{\phi}$ after a $1/4$ of the full period; see Fig. \ref{fig:14}}
\item{trajectories of type 'III' (which are the union of all previously mentioned cases) starting from the boundary set  $V(a_{0},\phi_{0})=0$ with $(\dot{a}_{0},\dot{\phi}_{0})=(0,0)$, after a $1/4$ of the full period they reach one inflection point at scale factor $a$, and $\phi=0$ with $(\dot{a},\dot{\phi})=(0,\pm a)$, after a $1/2$ of the period they reach the symmetrical point to the initial condition with respect to $\phi$--axis; see Fig. \ref{fig:15}}
\end{itemize}
Of course there are also non-symmetric periodic orbits present (see Fig. \ref{fig:16}) but they are not a subject of our consideration. In Table \ref{tab:1} we can find the periods of typical periodic orbits which we obtain by using modified multi shooting method \cite{Reithmeier:1991}. To illustrate the property of sensitive dependence on initial conditions we investigate the evolution of the separation vector in the phase space in the term of ``Lyapunov like'' principal exponent. The principal Lyapunov exponent as well as the distance of the separation vector for initial conditions' separation $\Delta a = 10^{-12}$, are illustrated in Fig. \ref{fig:17}. One can observe the existence of unstable periodic orbits in the model which should be treated as a strong evidence of chaos \cite{Cvitanovic:2002}.

\begin{table}[t]
\begin{center}
\begin{tabular}{|c|c|c|}
\hline
Type & $q^{1}_{0}=a_{0}$ & Period \\
\hline
I a &   1.90228463575087 & 8.21179541505428 \\
    &   4.07891339562106 & 9.68359976287486 \\
    &   6.14180160301955 & 10.3389393707077 \\
\hline
I b &   2.63614545223571 & 9.49181146895361 \\
    &   4.67384662462167 & 10.4432886891063 \\
    &   6.68822113801299 & 10.8883259280963 \\
\hline
II a &   3.08416398565973 & 9.12958397138681 \\
     &   5.16022385714180 & 10.06129848810491 \\
     &   7.19459890603674 & 10.55370414489915 \\
\hline
II b & 1.64890985955010 & 8.45628399932237 \\
     & 3.71764148794895 & 10.07328719135456 \\
     & 5.72295574169836 & 10.69967308122208 \\
\hline
III  & 2.47587020635594 &  17.99123630712047 \\
     & 2.75437747583008 &  19.56534435565128 \\
     & 2.94236929747301 &  20.15957644135879 \\
\hline
\end{tabular}
\end{center}
\caption{Unstable periodic trajectories.}
\label{tab:1}
\end{table}
					      
\section{Conclusions}

In the present paper the complex behavior of trajectories of FRW cosmological models with scalar field was investigated by means of geodesics of the Jacobi metric.

We pointed out the role of the singular set $\partial\mathcal{D}$ of degeneration of the Jacobi metric in detection of complexity of dynamical behavior. We find that this domain can be useful as a Poincar{\'e} surface. Moreover, the distribution of the intersection points as well as the existence of periodic orbits contain interesting information about the degree of complexity of its dynamics. Therefore all these investigations should be treated as a complementary description of chaotic behavior in a geometrical way.

We have demonstrated the complexity of dynamics in the sense of $(1)$ Poincar{\'e} sections, $(2)$ random distribution of intersection points, $(3)$ the existence of unstable periodic orbits and $(4)$ chaos in trajectories coding. All this evidence has rather mathematical sense because we prolong trajectories to the nonphysical domain $a<0$. The true sense of complexity following Motter and Letelier \cite{Motter:2002} is in the nonintegrability of its dynamics. In Ref. \cite{Maciejewski:2002,Maciejewski:2000,Maciejewski:2001} we have investigated the model under consideration in the framework of nonintegrability and we showed that they are non-integrable in the sense of nonexistence of meromorfic first integrals for the generic case of the model's parameters. To that end, the Yoshida, Ziglin, and strong methods of differential Galois group are used.

The presented approach offers some new possibilities of coding trajectories for which a well defined conception of Kolmogorov complexity can be also applied \cite{Kolmogorov:1965}. In 1963-1965 A.~N. Kolmogorov proposed to consider a measure of complexity in the framework of the general theory of algorithms. Let us consider model of dynamics in the form of mathematical object that has binary string $\alpha$ as its complete description. Then we can use the length of the shortest program in bits as a measure of the complexity of the object \cite{Li:1997}.

Let us now consider a cosmological model as a mathematical object that has coding dynamics in the form of binary string $\alpha$ as its complete description. Then for the purpose of quantifying complexity, the notion of Kolmogorov idea can be useful. The question, whether the FRW cosmological model is complex in the Kolmogorov sense, is open. In our opinion the geometrical language introduced here offers a new, interesting possibility of trajectories coding in the form of binary strings which makes the answering this question easier.

\end{document}